\documentclass[longauth]{aa} 
\usepackage{graphicx}
\usepackage{natbib}
\usepackage{amssymb}
\usepackage{amsmath}
\usepackage{amsfonts}
\bibpunct{(}{)}{;}{a}{}{,}
\usepackage{txfonts}

\usepackage{hyperref}
\usepackage[dvipsnames]{xcolor}

\begin{document}

   \title{The GRAVITY Young Stellar Object survey}
   \subtitle{VII. The inner dusty disks of T Tauri stars\thanks{GTO programs with run IDs: 0103.C-0347, 0102.C-0408, 0101.C-0311, 0100.C-0278, and 099.C-0667}}
      \author{The GRAVITY Collaboration: K. Perraut\inst{1} \and L. Labadie\inst{2} \and  J. Bouvier\inst{1} \and F. M\'enard\inst{1} \and L. Klarmann\inst{3} \and  C. Dougados\inst{1} \and M. Benisty\inst{1,4} \and J.-P. Berger\inst{1} \and Y.-I. Bouarour\inst{5,11} \and W. Brandner\inst{3} \and A. Caratti o Garatti\inst{5,15} \and P. Caselli\inst{6} \and P.T. de Zeeuw\inst{6,9} \and R. Garcia-Lopez\inst{3,5,11} \and T. Henning\inst{3} \and J. Sanchez-Bermudez\inst{7,3} \and A. Sousa\inst{1} \and E. van Dishoeck\inst{6,9} \and E. Al\'ecian\inst{1} \and A. Amorim\inst{12,13} \and Y. Cl\'enet\inst{8} \and R. Davies\inst{6} \and A. Drescher\inst{6} \and G. Duvert\inst{1} \and A. Eckart\inst{2,10} \and F. Eisenhauer\inst{6} \and N.M. F\"orster-Schreiber\inst{6} \and P. Garcia\inst{12,14} \and E. Gendron\inst{8} \and R. Genzel\inst{6} \and S. Gillessen\inst{6} \and R. Grellmann\inst{2} \and G. Hei$\ss$el\inst{8} \and S. Hippler\inst{3} \and M. Horrobin\inst{2} \and Z. Hubert\inst{1} \and L. Jocou\inst{1} \and P. Kervella\inst{8} \and S. Lacour\inst{8} \and V. Lapeyr\`ere\inst{8} \and J.-B. Le Bouquin\inst{1} \and P. L\'ena\inst{8} \and D. Lutz\inst{6} \and T. Ott\inst{6} \and T. Paumard\inst{8} \and G. Perrin\inst{8} \and S. Scheithauer\inst{3} \and J. Shangguan\inst{6} \and T. Shimizu\inst{6} \and J. Stadler\inst{6} \and O. Straub\inst{6} \and C. Straubmeier\inst{2} \and E. Sturm\inst{6} \and L. Tacconi\inst{6} \and F. Vincent\inst{8} \and S. von Fellenberg\inst{6} \and F. Widmann\inst{6}
      }

  \institute{
  Univ. Grenoble Alpes, CNRS, IPAG, 38000 Grenoble, France
\and
I. Physikalisches Institut, Universit\"at zu K\"oln, Z\"ulpicher Strasse 77, 50937, K\"oln, Germany
\and
Max Planck Institute for Astronomy, K\"onigstuhl 17,
69117 Heidelberg, Germany
\and
Unidad Mixta Internacional Franco-Chilena de Astronom\'ia (CNRS UMI 3386), Departamento de Astronom\'ia, Universidad de Chile, Camino El Observatorio 1515, Las Condes, Santiago, Chile     
\and
Dublin Institute for Advanced Studies, 31 Fitzwilliam Place, D02\,XF86 Dublin, Ireland
\and
Max Planck Institute for Extraterrestrial Physics, Giessenbachstrasse, 85741 Garching bei M\"{u}nchen, Germany
\and
Instituto de Astronom\'ia, Universidad Nacional Aut\'onoma de M\'exico, Apdo. Postal 70264, Ciudad de M\'exico 04510, Mexico
\and
LESIA, Observatoire de Paris, PSL Research University, CNRS, Sorbonne Universit\'es, UPMC Univ. Paris 06, Univ. Paris Diderot, Sorbonne Paris Cit\'e, France
\and
Leiden Observatory, Leiden University, Postbus 9513, 2300 RA
Leiden, The Netherlands
\and
Max-Planck-Institute for Radio Astronomy, Auf dem Hügel 69, 53121 Bonn, Germany
\and
School of Physics, University College Dublin, Belfield, Dublin 4, Ireland
\and
CENTRA, Centro de Astrof\'{\i}sica e Gravita\c{c}\~{a}o, Instituto Superior T\'{e}cnico, Avenida Rovisco Pais 1, 1049 Lisboa, Portugal
\and
Universidade de Lisboa, Faculdade de Ciências, Campo Grande, 1749-016 Lisboa, Portugal
\and
Universidade do Porto, Faculdade de Engenharia, Rua Dr. Roberto Frias, 4200-465 Porto, Portugal
\and
INAF -- Osservatorio Astronomico di Capodimonte, via Moiariello 16, 80131 Napoli, Italy
\\
     Email: karine.perraut@univ-grenoble-alpes.fr   
}
   \date{Received ; accepted }

   \date{Received ; accepted }

 
  \abstract
   {T Tauri stars are surrounded by dust and gas disks. As material reservoirs from which matter is accreted onto the central star and planets are built, these protoplanetary disks play a central role in star and planet formation.
   }
   {We aim at spatially resolving at sub-astronomical unit (sub-au) scales the innermost regions of the protoplanetary disks around a sample of T Tauri stars to better understand their morphology and composition.
   }
   {Thanks to the sensitivity and the better spatial frequency coverage of the GRAVITY instrument of the Very Large Telescope Interferometer (VLTI), we extended our homogeneous data set of 27 Herbig stars and collected near-infrared K-band interferometric observations of 17 T Tauri stars, spanning effective temperatures and luminosities in the ranges of $\sim$~4000--6000~K and $\sim$~0.4--10~$L_\odot$, respectively. We focus on the continuum emission and develop semi-physical geometrical models to fit the interferometric data and search for trends between the properties of the disk and the central star.
   }
   {As for those of their more massive counterparts, the Herbig Ae/Be stars, the best-fit models of the inner rim of the T Tauri disks correspond to wide rings. The GRAVITY measurements extend the Radius-luminosity relation toward the smallest luminosities (0.4-10~L$_\odot$). As observed previously, in this range of luminosities, the $R$~$\propto$~$L^{1/2}$ trend line is no longer valid, and the K-band sizes measured with GRAVITY appear to be larger than the predicted sizes derived from sublimation radius computation. We do not see a clear correlation between the K-band half-flux radius and the mass accretion rate onto the central star. Besides, having magnetic truncation radii in agreement with the K-band GRAVITY sizes would require magnetic fields as strong as a few kG, which should have been detected, suggesting that accretion is not the main process governing the location of the half-flux radius of the inner dusty disk. The GRAVITY measurements agree with models that take into account the scattered light, which could be as important as thermal emission in the K band for these cool stars. The N-to-K band size ratio may be a proxy for disentangling disks with silicate features in emission from disks with weak and/or in absorption silicate features (i.e., disks with depleted inner regions and/or with large gaps). 
   The GRAVITY data also provide inclinations and position angles of the inner disks. When compared to those of the outer disks derived from ALMA images of nine objects of our sample, we detect clear misalignments between both disks for four objects. 
      }
   {The combination of improved data quality with a
  significant and homogeneous sample of young stellar objects allows us to revisit the pioneering works done on the protoplanetary disks by K-band interferometry and to test inner disk physics such as the inner rim morphology and location.
   }

   \keywords{stars: formation -- stars: circumstellar matter  -- Infrared: ISM -- Instrumentation: interferometers -- Techniques: high angular resolution -- Techniques: interferometric}

    \authorrunning {K. Perraut et al.}
    \titlerunning{The GRAVITY YSO survey - The inner dusty disk of T Tauri stars}
   \maketitle
%

\section{Introduction}

\begin{figure*}[t]
    \centering
    \includegraphics[width=13cm]{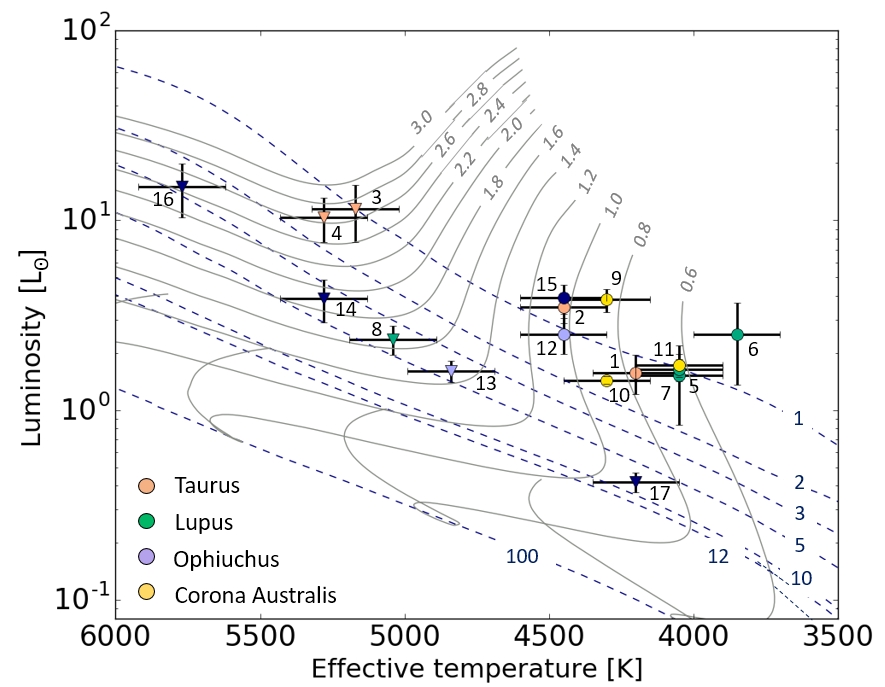}
    \caption{Our T Tauri sample set in the Hertzsprung-Russell diagram. The colors of the symbols denote the star forming regions. Dashed blue lines denote the isochrones for 1, 2, 3, 5, 10, 12, and 100 Ma, and solid gray lines the iso-mass tracks from 0.6 to 3~M$_\odot$. The tracks are PMS PISA tracks \citep{PISA} for solar abundances ($Z$~=~0.02 and $Y$~=~0.288), a solar-calibrated mixing length ($\alpha$~=~1.68), and an initial deuterium abundance of 2$\times$10$^{-5}$. The circles denote the fully convective stars, while the triangles denote the stars with a radiative core and a convective envelope.}
    \label{fig:HRD}
\end{figure*}
During their early pre-main-sequence (PMS) phase, the 1-2~M$_\odot$ young stars, namely the classical T Tauri stars, are surrounded by protoplanetary disks. These disks play a crucial role in stellar and planetary formation as they are reservoirs from which both stars and planets accrete material. The structure and the evolution of the inner disk regions of T Tauri stars impact both the accretion onto the star and the processing of dust grains in the terrestrial planet-forming region (0.1--1 astronomical unit (au) region). Exploring the inner disk structure is therefore of great importance to understand the star-disk interactions and the mechanisms of planet formation. The stellar magnetic field truncates the inner edge of the disk at typically a few stellar radii from the stellar surface. Gas from the disk is funneled along the magnetic field lines, down to the stellar surface \citep{Bouvier2007,Hartmann2016}. Dust sublimation is the commonly assumed mechanism that sets the inner rim of the dusty disk. Just beyond the sublimation front, a vortex can form at the dead zone boundary and create a dust trap where dust grains can grow to pebbles and planetesimals \citep{Regaly2012,Flock_2016,Flock_2017}. The processes of disk evolution and planet formation are thus intrinsically linked.

At the typical distance of the closest low-mass star forming regions ($\sim$~140~pc), probing the star-disk interaction region and the inner rim of the protoplaneraty disks requires an angular resolution of a millisecond of arc (mas) or less, which is out of reach for the existing largest telescopes. Thus, for a long time these phenomena were mostly investigated through indirect techniques (photometry, spectroscopy, and spectro-polarimetry), until optical long-baseline interferometry allowed us to reach these scales \citep{MillanGabet,MaccAS205,Eisner2007,Eisner2010}. These pioneering works in the near-infrared range provided new insights into the morphological properties of the inner disks of young stellar objects \citep[YSOs;][]{dullemond10,kraus2015}. In particular, the size-luminosity relation between the measured inner radius of the disk and the luminosity of the central star \citep{Monnier2005} strongly suggests the presence of a directly illuminated rim at the dust sublimation radius \citep{Natta2001}. These interferometric measurements coupled with the (spectro-)photometric ones have allowed several disk models to be put to the test, particularly regarding the inner rim structure, for example, a puffed-up rim \citep{isella}, a wedge-shaped rim \citep{Tannikurlam2008}, or a self-consistently calculated rim structure that includes both rim shapes \citep{Kama2009}. The structure and the composition of these regions are still a matter of debate.

\begin{table*}[t]
        \caption{Properties of the stars of the T Tauri sample.}
    \centering
    \begin{tabular}{cccccccccc}
    \hline
    \# & Name & Sp. Type & $T_{\rm eff}$ & $L$ & $d$  & $M$  & age  & $\dot{M}_{acc}$
    & References\\
     & &  & [K] & [L$_\odot$] & [pc] & [M$_\odot$] & [Ma] & [M$_\odot$/yr] 
     & \\
    \hline
        1 & DG Tau & K6 & 4200 & 1.58~$\pm$~0.37 & 125.2~$\pm$~2.3 & 0.8~$\pm$~0.2 & 1-2 & 4.6-74 10$^{-8}$ 
        & \cite{MaccDGTau}\\
        2 & DR Tau & K5 & 4450 & 3.48~$\pm$~0.61 & 193.0~$\pm$~1.9 & 1.0~$\pm$~0.2 & $\leq$~1 & 5.2 10$^{-8}$ 
        & \cite{MaccDRTau}\\
        3 & RY Tau & K1 & 5170 & 11.5~$\pm$~3.8 & 138.2~$\pm$~4.8 & 2.7~$\pm$~0.3 & 1-2 & 6-9 10$^{-9}$ 
        & \cite{MaccRYTauTTau}\\
        4 & T Tau N & K0 & 5280 & 10.4~$\pm$~2.7 & 145.1~$\pm$~1.0 & 2.6~$\pm$~0.3 & 1-2 & 3-5 10$^{-8}$
        &\cite{MaccRYTauTTau}\\
        5 & GQ Lup & K7 & 4050 & 1.52~$\pm$~0.68$^{(a)}$ & 154.1~$\pm$~0.9 & 0.6~$\pm$~0.2 & 1-2 & 2 10$^{-8}$ 
        & \cite{MaccGQIMRYLup}\\
        6 & IM Lup & M0 & 3850 & 2.51~$\pm$~1.15$^{(a)}$ & 155.8~$\pm$~1.0 & 0.4~$\pm$~0.2 & $\leq$~1 & 6.3 10$^{-9}$ 
        & \cite{MaccGQIMRYLup}\\
        7 & RU Lup & K7 & 4050 & 1.64~$\pm$~0.22 & 157.5~$\pm$~1.3 & 0.6~$\pm$~0.2 & 1-2 & 4-11 10$^{-8}$ 
        &\cite{MaccGQIMRYLup}\\
        8 & RY Lup & K2 & 5040 & 2.36~$\pm$~0.41$^{(b)}$ & 153.5~$\pm$~1.8 & 1.7~$\pm$~0.2 & 3-6 & 6.3 10$^{-9}$ 
        & \cite{MaccGQIMRYLup}\\
        9 & S CrA N & K5-6 & 4300  & 3.82~$\pm$~0.54 & 160.5~$\pm$~2.2 & 0.9~$\pm$~0.2 & $\leq$~1 & 10$^{-7}$ 
        & \cite{MaccVVCrA}\\
        10 & S CrA S & K5-6 & 4300 & 1.44~$\pm$~0.07 & 147.4~$\pm$~2.1 & 0.9~$\pm$~0.2 & 1-2 & 3 10$^{-8}$ 
        &\cite{MaccVVCrA} \\
        11 & VV CrA SW$^{(*)}$ & K7 & 4050 & 1.73~$\pm$~0.24 & 156.6~$\pm$~1.2 & 0.6~$\pm$~0.2 & 1-2 & 2-3 10$^{-8}$ 
        &\cite{MaccVVCrA}\\
        12 & V2062 Oph & K2-3 & 4840 & 1.61~$\pm$~0.21$^{(c)}$ & 146.3~$\pm$~0.6 & 1.5~$\pm$~0.2 & 3-6 & 6.3 10$^{-9}$ 
        &\cite{MaccDoAr44TWHya} \\
        13 & V2129 Oph & K5 & 4450 & 2.51~$\pm$~0.54 & 131.9~$\pm$~0.5 & 1.0~$\pm$~0.2 & 1-2 & 1.5 10$^{-9}$ 
        & \cite{MaccV2129Oph}\\
        14 & AS 205 N & K0 & 5280 & 3.89~$\pm$~0.98 & 132.1~$\pm$~1.6 & 1.9~$\pm$~0.2 & 2-6 & 7.2 10$^{-7}$ 
        &\cite{MaccAS205} \\
        15 & AS 353 & K5 & 4450 & 3.90~$\pm$~0.67$^{(d)}$ & 399.6~$\pm$~4.2 & 1.0~$\pm$~0.2 & $\leq$~1 & 7.9 10$^{-8}$ 
        &\cite{MaccAS353} \\
        16 & DI Cha & G2 & 5770 & 15.1~$\pm$~4.8 & 189.0~$\pm$~0.6 & 2.4~$\pm$~0.4 & 2-5 & 5.98 10$^{-8}$ 
        &\cite{MaccDICha}\\
        17 & TW Hya & K6 & 4200 & 0.41~$\pm$~0.05 & 60.14~$\pm$~0.07 & 0.9~$\pm$~0.1 & 4-12 & 2 10$^{-9}$ 
        &\cite{MaccDoAr44TWHya}\\
        \hline
    \end{tabular}
    \label{tab:param_fond}
    \tablefoot{The effective temperatures are derived from the spectral types. Unless noted otherwise, the intrinsic luminosities are taken from \citet{Varga2018} and rescaled with distances from Gaia EDR3 \citep{GaiaDR3}. The masses and the ages are derived from the evolutionary PISA tracks \citep{PISA} displayed in Fig.~\ref{fig:HRD}. The accretion rates are taken from the references given in the last column of the Table.
    \noindent $^{(a)}$ L from \cite{MaccGQIMRYLup}; $^{(b)}$ L from \citet{Bouarour2020} with an error of 15\%; $^{(c)}$ L from \citet{Bouvier2020b}; $^{(d)}$ L from \citet{Hamann1992} with an error of 15\%; $^{(e)}$ $L_{\rm acc}$ from \citet{Eisner2010}; $^{(*)}$ This component is also called VV~CrA~South or VV~CrA~A.}
\end{table*}

Due to the limited sensitivity of the first-generation interferometric instruments, near-infrared observations of YSOs initially focused on the Herbig stars that are hotter and brighter than T Tauri stars. As a consequence, T Tauri's circumstellar disks have an inner rim closer to their central star, and resolving it with 100 m baselines becomes more challenging. About 25 T Tauri stars were observed by \citet{Eisner2007,Eisner2010,Eisner2014} and \citet{Akeson2005} with the two-telescope Keck Interferometer (KI) in the K band; a few of them were observed with three telescopes and the Very Large Telescope Interferometer (VLTI)/AMBER instrument \citep{vural12,Olofsson2011}; the first survey in the H band, which included 21 T Tauri stars, was led by \citet{Anthonioz2015} with the PIONIER instrument \citep{lebouquin}, which combines four telescopes and thus simultaneously probes different baseline orientations and measures closure phases, permitting the detection of departures from centro-symmetry. The authors observed 18 T Tauris, resolved 13 disks, and for most of them detected a significant contribution from an extended (e.g., larger than 3 au at 150~pc) contribution of light from the disk, which is compatible with scattered light as proposed previously by \citet{Pinte2008}.

In the mid-infrared range, the two-telescope instrument MIDI \citep{Leinert2003} observed 82 protoplanetary disks, among which 45 were around T Tauri stars. Based on this large data set, which spans a large range of stellar luminosities (0.2~L$_\odot$~$\leq$~$L_*$~$\leq$~4 10$^4$~L$_\odot$), two statistical studies were conducted \citep{Menu2015A&A...581A.107M, Varga2018}. From the mid-infrared size–luminosity relation, these studies show that the size of the emitting region does not simply scale as $L_*^{1/2}$. The disks around T Tauri stars appear generally colder and more extended with respect to the stellar luminosity than those around the Herbig/Ae stars of their sample. As the silicate feature is generally weaker at au or sub-au scales than in the outer parts of the disks, the dust might be substantially more processed in the innermost regions.

Here we aim at investigating the innermost regions of the protoplanetary disks of T Tauri stars with a homogeneous approach. 
We study the location of the dust emission in the K band, how it scales with the central star properties (such as luminosity, mass, accretion rate), how it compares with the size measured in the mid-infrared range, and whether an extended component is detected as in the PIONIER observations in the H band. To do so, we have taken advantage of the drastic improvement in sensitivity and stability offered by the GRAVITY instrument in the near-infrared K band \citep{GRAVITY}. Able to track fringes on stars as faint as Kmag~=~10 (Kmag~=~7, respectively) when combining the four Unit Telescopes (Auxiliary Telescopes, respectively) of the VLTI, this instrument gives access to a significant number of T Tauri stars that can be probed along six different interferometric baselines at the same time. Since its installation at the combined focus of the VLTI at Cerro Paranal in 2016, we have already observed a large number of YSOs within the framework of the Guaranteed Time Observations (GTO) YSO program. As first results, we directly probed both the accretion flows within the magnetosphere of classical T Tauri stars \citep{Bouvier2020a,GarciaLopez2020} and the K-band emission at the inner rim of protoplanetary disks of Herbig Ae/Be stars (\cite{PaperI}; hereafter GC19). 


In the present paper we report on new K-band continuum observations of 17 less massive classical T Tauri stars observed with GRAVITY, enlarging our sample to 44 YSOs. The paper is organized as follows: Our sample is described in Sect. 2; GRAVITY observations and data are detailed in Sect. 3; we present our geometrical model in Sect. 4; and the results are given in Sect. 5 and discussed in Sect. 6.

\section{Sample}

In GC19, we studied the thermal emission around 27 Herbig Ae/Be stars with spectral types ranging from B2 to F8 (e.g., effective temperatures varying between 6375~K and 17500~K), luminosities between 1 to 10$^4$~L$_\odot$, masses between 1.4 and 12.2~M$_\odot$, and an age coverage of 0.04--14.5~Ma. In the present study, we extend our young star sample toward the T Tauri stars with 17 new targets with spectral types ranging from G2 to M0. We selected our targets to be as bright as Kmag~=~7-8 in agreement with the GRAVITY sensitivity with the Auxiliary Telescopes. We use the spectral types from \citet{pecaut13} to derive the effective temperatures of our targets. The median effective temperature of our sample equals 4450~K and all our targets have an effective temperature lower than 5800~K. For stars observed by \citet{Varga2018}, we used the intrinsic luminosities provided by their Spectral Energy Distribution (SED) fits and their extinction estimates, else we used the values provided in the literature (Table~\ref{tab:param_fond}). We rescaled all luminosities to the Gaia EDR3 distance and considered errors on the parallaxes equal to 1.3 times the errors given in the catalog, as recommended in \citet{GaiaError} for sources brighter than Gmag~=~13, which is the case of our targets. The median luminosity of our T Tauri sample is 2.5~L$_\odot$, with 14 among 17 stars with a luminosity smaller than 10~L$_\odot$. We thus explore another part of the Hertzsprung-Russell diagram and, using the PISA tracks for PMS stars \citep{PISA}, we can derive the masses and the ages of our targets (Fig.~\ref{fig:HRD}): the mass of 11 out of 17 stars is lower than 1~M$_\odot$, and 12 out of 17 stars are younger than 2~Ma. Apart from TW~Hya, the less massive stars are still on the Hayashi tracks, and thus fully convective (DG ~Tau, DR~Tau, GQ~Lup, IM~Lup, RU~Lup, S~CrA~N, S~CrA~S, VV~CrA~SW, V2129~Oph, AS~353); the seven other targets (RY~Tau, T~Tau~N, RY~Lup, V2062~Oph, AS~205~N, DI~Cha, TW~Hya) have reached the radiative tracks and have thus a radiative core with a convective envelope. Regarding the accretion properties, 11 among 17 stars have an accretion rate higher than 10$^{-8}$~M$_\odot$.yr$^{-1}$.

Among our sample, seven stars are part of the PIONIER survey (TW~Hya, GQ~Lup, RU~Lup, RY~Lup, AS~205~N, V2129~Oph, S~CrA~N); 13 stars are in the MIDI atlas (TW~Hya, DG~Tau, DR~Tau, RY~Tau, T~Tau~N, RU~Lup, S~CrA~N, S~CrA~S, VV~CrA~SW, V2062~Oph, V2129~Oph, AS~205~N, DI~Cha); and six have been previously observed with the KI (DG~Tau, DR~Tau, RY~Tau, T~Tau~N, AS~205~N, AS~353).

\section{GRAVITY observations and data}

We observed all targets with the GRAVITY instrument, which combines the light from four telescopes, either the 8~m Unit Telescopes (UT) or the 1.8~m Auxiliary Telescopes (AT). With GRAVITY, the interferometric fringes on the six baselines (B) are recorded simultaneously on the fringe tracker (FT) and on the scientific instrument (SC) detectors. Working at a high frame rate (i.e., 300 or 900~Hz) to freeze the atmospheric perturbations, the FT operates at low spectral resolution with six spectral channels over the K band \citep{Lacour2019A&A...624A..99L}. When the fringes are locked by the FT, the SC can record the interferometric observables during long exposure times (as long as a few tens of seconds) and can provide a spectral resolution as high as R~$\sim$~4000 over the whole K band. For each target of our program, we recorded several 5-minute-long files on the object itself, and interleaved these observations with observations of interferometric calibrators. The calibrators were carefully selected to be single stars, close to the target, of similar magnitude as the target, and unresolved (small angular diameters). The log of the observations is given in Appendix A. 


To carry out a homogeneous study of the inner dust rim in the K-band continuum for our sample of 44 YSOs, we followed the same approach as in GC19 and focused on the FT data, which are the most robust against the effects of the atmospheric turbulence. We reduced all our data with the GRAVITY data reduction pipeline \citep{DRS}. The atmospheric transfer function for each night was calibrated using the calibrator observations. For each file on a target, we obtained six squared visibilities and four closure phases for five spectral channels. The first channel (e.g., at the lowest wavelength) should be discarded since it may be affected by the metrology laser working at 1.908 µm and by the strong absorption lines of the atmospheric transmission. The SC observations for the targets exhibiting a hydrogen Br$\gamma$ line in emission (marked in bold in Table~\ref{tab:obs}.1) have already been published for S~CrA \citep{Garcia-Lopez2017}, TW~Hya \citep{GarciaLopez2020}, and V2062~Oph \citep{Bouvier2020b}, or will be analyzed in a forthcoming paper.
 

All the calibrated data of the FT are presented in Appendix B along with the corresponding {\it (u, v)} spatial frequency planes. For most of our targets, our GRAVITY observations span a spatial frequency range between 20~M$\lambda$ and 60~M$\lambda$, leading to an angular resolution of $\lambda$/2B of about 1.7 mas at maximum, which corresponds to about 0.25~au at a distance of 140~pc. The variation in the visibility with spatial frequency could be fitted with geometrical models to derive the extent of the environment, while the closure phase is related to the asymmetry of the environment \citep{Haniff2007}. In our sample, most of our targets appear to be partially resolved in the continuum with minimum squared visibilities ranging from 0.4 to 0.9. A few of them exhibit squared visibilities as small as 0.2-0.3 (DG~Tau, RY~Tau, T~Tau N, AS~205~N, and VV~CrA SW) or a possible plateau of the squared visibilities at the longest baselines (RU~Lup), which could indicate that the dusty environments are fully resolved (see Sect. 5.2).
Regarding the closure phase observables, 2/3 of our targets exhibit closure phases consistent with 0$^\circ$, which indicates centro-symmetric objects at our resolution; for the other targets, most of the closure phase signals are smaller than 5$^\circ$ (only AS~205~N has a closure phase slightly larger than 5$^\circ$). The targets exhibiting the largest closure phases are also among the most resolved targets (with minimum squared visibilities below 0.4), noting however that two other fully resolved targets (DG~Tau and T~Tau N) have small closure phases. 

\begin{table*}[t]
        \caption{Best-fit parameters for the centro-symmetric Gaussian ring model with 1~$\sigma$ error bars. The $\chi^2_r$ computation includes both visibilities and closure phases. For the sake of comparison, the $\chi^2_r$ values in parentheses correspond to the centro-symmetric model without the halo contribution. See Section~\ref{sec:best-fit} for details. 
        }
    \centering
    \begin{tabular}{cccccccccccc}
    \hline
    \# & Name & $\cos i$ & $i$ & PA & $f_c$ & $f_h$ & \multicolumn{2}{c}{Half-flux radius $a$} & $w$& $\chi^2_r$ &  \\
      & -- & -- & [$^\circ$] & [$^\circ$] & [\%] & [\%] & [mas] & [au] & -- & -- & 
    \\
    \hline
        1 & DG Tau & 0.65~$\pm$~0.04 & 49~$\pm$~4 & 143~$\pm$~12 & 45~$\pm$~5 & 31~$\pm$~2 & 1.0~$\pm$~0.1 & 
        0.125~$\pm$~0.01 & 0.997$^{+0.003}_{-0.08}$ & 1.94 { (4.56)} &  
        \\ [1ex]
        2 & DR Tau & 0.95~$\pm$~0.07 & 18$^{+10}_{-18}$ & 14~$\pm$~41 & 90~$\pm$~5 & 0~$\pm$~2 & 0.48$^{+0.07}_{-0.06}$ & 
        0.092$^{+0.014}_{-0.012}$ & 0.17$^{+0.64}_{-0.12}$ & 2.97 { (2.96)}& 
        \\ [1ex]
        3 & RY Tau & 0.50~$\pm$~0.01 & 60~$\pm$~1 & 8~$\pm$~1 & 50~$\pm$~5 & 7~$\pm$~2 & 2.45$^{+0.24}_{-0.22}$ & 
        0.34~$\pm$~0.03 & 0.99$^{+0.01}_{-0.05}$ & 4.77 { (5.09)}& \\ [1ex]
        4 & T Tau N & 0.82~$\pm$~0.04 & 35~$\pm$~4 & 93~$\pm$~9 & 31~$\pm$~4 & 9~$\pm$~2 & 1.74~$\pm$~0.08 & 
        0.25~$\pm$~0.01 & 0.10$^{+0.02}_{-0.01}$ & 2.05 { (3.63)}& 
        \\ [1ex]
        5 & GQ Lup & 0.93~$\pm$~0.03 & 22~$\pm$~6 & 180~$\pm$~3 & 40~$\pm$~20 & 7~$\pm$~2 & 0.81~$\pm$~0.12 & 
        0.13~$\pm$~0.02 & 0.11$^{+0.33}_{-0.09}$ & 3.14 { (5.33)}& \\ [1ex]
        6 & IM Lup & 0.51~$\pm$~0.05 & 59~$\pm$~4 & 139~$\pm$~3 & 40~$\pm$~20 & 5~$\pm$~2 & 1.00$^{+0.38}_{-0.28}$ & 
        0.16$^{+0.06}_{-0.04}$& 0.30$^{+0.63}_{-0.26}$ & 2.64 { (4.02)}& \\ [1ex]
        7 & RU Lup &0.96~$\pm$~0.03 & 16$^{+6}_{-8}$ & 99~$\pm$~31 & 30~$\pm$~10 & 12~$\pm$~3 & 1.32$^{+0.42}_{-0.32}$ & 
        0.21$^{+0.07}_{-0.05}$ & 0.71$^{+0.29}_{-0.61}$ & 2.14 { (2.87)} & 
        \\ [1ex]
        8 & RY Lup &0.60~$\pm$~0.07 & 53~$\pm$~5 & 73~$\pm$~2 & 25~$\pm$~2 & 10~$\pm$~2 & 1.00~$\pm$~0.07 & 
        0.16~$\pm$~0.01 & 0.99$^{+0.01}_{-0.08}$ & 1.09 { (2.10)}& 
        \\ [1ex]
        9 & S CrA N & 0.89~$\pm$~0.02 & 27~$\pm$~3 & 1~$\pm$~6 & 50~$\pm$~10 & 6~$\pm$~2 & 1.05$^{+0.15}_{-0.14}$ & 
        0.17~$\pm$~0.02 & 0.49$^{+0.45}_{-0.38}$ & 1.13 { (1.84)}& \\ [1ex]
        10 & S CrA S & 0.86~$\pm$~0.04 & 31~$\pm$~5 & 176~$\pm$~8 & 40~$\pm$~10 & 9~$\pm$~2 & 0.83$^{+0.08}_{-0.07}$ & 
        0.12~$\pm$~0.01 & 0.15$^{+0.59}_{-0.13}$ & 0.82 { (4.38)}& \\ [1ex]
        11 & VV CrA SW & 0.85~$\pm$~0.02 & 32~$\pm$~3 & 91~$\pm$~6 & 84~$\pm$~4 & 14~$\pm$~3 & 1.00~$\pm$~0.05 & 
        0.157~$\pm$~0.007 & 0.994$^{+0.005}_{-0.04}$ & 5.24 { (9.46)}& 
        \\ [1ex]
        12 & V2062 Oph & 0.85~$\pm$~0.03 & 32~$\pm$~4 & 137~$\pm$~4 & 21~$\pm$~12 & 6~$\pm$~2 & 1.07$^{+0.13}_{-0.12}$ & 
        0.16~$\pm$~0.02 & 0.95$^{+0.05}_{-0.42}$ & 0.67 { (1.09)}& \\ [1ex]
        13 & V2129 Oph & 0.66~$\pm$~0.06 & 49~$\pm$~5 & 77~$\pm$~7 & 28~$\pm$~6 & 6~$\pm$~2 & 0.79$^{+0.14}_{-0.13}$ & 
        0.10~$\pm$~0.02 & 0.98$^{+0.02}_{-0.14}$ & 1.10 { (1.30)}& 
        \\ [1ex]
        14 & AS 205 N & 0.72~$\pm$~0.02 & 44~$\pm$~2 & 90~$\pm$~1 & 50~$\pm$~10 & 9~$\pm$~2 & 1.17$^{+0.09}_{-0.07}$ & 
        0.16~$\pm$~0.01 & 0.125$^{+0.06}_{-0.04}$ & 3.92 { (5.92)}&\\ [1ex]
        15 & AS 353 & 0.76~$\pm$~0.02 & 41~$\pm$~2 & 173~$\pm$~3 & 50~$\pm$~20 & 9~$\pm$~2 & 0.71$^{+0.12}_{-0.11}$ & 
        0.28$^{+0.05}_{-0.04}$ & 0.98$^{+0.02}_{-0.11}$ & 0.62 { (2.70)}&\\ [1ex]
        16 & DI Cha & 0.86~$\pm$~0.02 & 31~$\pm$~3 & 42~$\pm$~5 & 60~$\pm$~10 & 6~$\pm$~2 & 0.95~$\pm$~0.05 & 
        0.18~$\pm$~0.01& 0.14$^{+0.06}_{-0.04}$ & 3.16 { (4.02)}& \\ [1ex]
        17 & TW Hya & 0.97~$\pm$~0.03 & 14$^{+6}_{-14}$ & 130~$\pm$~32 & 12~$\pm$~2 & 2~$\pm$~2 & 0.69~$\pm$~0.05 & 
        0.042~$\pm$~0.003 & 0.89$^{+0.10}_{-0.35}$ & 0.16 { (0.26)}& 
        \\ [1ex]
        \hline
    \end{tabular}
    \label{tab:fits_ring}
\end{table*}

\section{Geometrical model}
\label{sect:model}

The visibility variation with the spatial frequency allows us to determine the geometrical properties of the circumstellar environment through model fitting. We followed the same geometrical approach as in GC19; we fitted our visibility curves with a geometrical model consisting of a point-like source and a circumstellar environment composed of two components, a ring and an extended component (or halo). Because of a possible degeneracy between the halo flux contribution and the radial brightness distribution, we impose a Gaussian radial brightness distribution for the ring to ensure that, for all targets of our sample, the halo flux contribution is derived in a homogeneous way. The visibility as a function of spatial frequency and wavelength is thus given by:
\begin{equation}
    V (u,v,\lambda) = \frac{f_s (\lambda_0/\lambda)^{\rm k_s} + f_c (\lambda_0/\lambda)^{\rm k_c} V_c (u,v)}{(f_s + f_h) (\lambda_0/\lambda)^{\rm k_s} + f_c (\lambda_0/\lambda)^{\rm k_c}},
\end{equation}
with $f_s$, $f_c$, $f_h$ the fractional flux contributions of the star, the disk, and the halo, respectively; $V_c$ the visibility of the ring; $\lambda_0$~=~2.15~$\mu$m the wavelength of the central spectral channel of the FT; $\rm k_s$ and $\rm k_c$ the spectral indices of the star and of the disk, respectively.

As mentioned in \cite{lazareff17}, the flux contribution and the size of the ring can be degenerate when the object is only partially resolved, which is the case for the majority of our objects. This implies that a compact ring with a large fractional flux or an extended ring with a small fractional flux would contribute similarly to the total visibility.
Thus, as in GC19, for each target, we used the near-infrared excess in the K band derived from a fit of the Spectral Energy Distribution (SED) as a starting value for the fractional flux of the ring $f_c$, when starting the fit of our interferometric data. We used the SEDs from \cite{Varga2018}, which allows us to take into account the variability of the source in the determination of the errors of the derived parameters. The near-infrared excess is only used as a starting value, and $f_c$ is kept free during the fitting process. For two targets (TW~Hya and V2129~Oph), the excesses in the K band have been accurately determined by high resolution spectroscopy \citep{MaccDoAr44TWHya,Sousa2021}, so we adopted these values and their errors to derive the fitted parameters and their errors. For all objects, we checked that there is no additional local minima in the convergence process.

We refer to GC19 for the determination of the spectral indices, and for the details of the fitting processes (see Sect. 4.4 in GC19). For the analytical expression of the visibility, we refer to Table 5 of \cite{lazareff17} and consider hereafter:
\vskip 0.05cm
{\it The centro-symmetric ring model.} This model has seven free parameters: the fractional flux contributions of the ring, $f_c$, and of the halo, $f_h$ (the star flux contribution being given by $f_s$~=~1~-~$f_c$~-~$f_h$), the inclination $i$ and the position angle $PA$, the half-flux radius $a$, the width-to-radius ratio $w$, and the spectral index of the ring $k_c$. 
\vskip 0.05cm
{\it The azimuthally modulated ring model.} An azimuthal modulation of the brightness profile can be added to mimic azimuthal variations in the surface brightness in the ring due to the inclination effect, density variation, or asymmetric scattering phase function. Such a model produces non-zero closure phases, and has two additional free parameters, $c_1$ and $s_1$ \citep[i.e., the cosine and sine angular modulation of first order; see Eq. (8) in][]{lazareff17}.

As in GC19, for all fits, we impose floor values to the error estimates coming from the reduction pipeline of respectively 2\% for the squared visibilities and 1$^\circ$ for the closure phases.

\section{Results}

\subsection{Best fits of the interferometric data}\label{sec:best-fit}

 Since the measured closure phases are small ($\leq\pm$\,5$^\circ$), we first fitted all our data sets with the centro-symmetric ring model, { with and without a halo contribution. When removing the halo contribution, the $\chi^2_r$ values are worse (up to a factor of 2.3 in the extreme case of DG~Tau; see the $\chi^2_r$ comparison in the last column of Table~\ref{tab:fits_ring}). Hereafter, we thus consider the best fits of our K-band data that correspond to the centro-symmetric ring model with a halo. The fit parameters are} given in Table~\ref{tab:fits_ring}. The residuals of these fits are displayed in Figs. B.1-B.4 as blue symbols at the bottom of the visibility and closure phase curves. {For these best-fit models, the $\chi^2_r$ values, which are computed on both visibilities and closure phases,} are on the order of 3 or less, except for the targets for which the closure phase signals clearly depart from 0: RY~Tau, VV~CrA~SW, and AS~205~N. Even for these three targets, the $\chi^2_r$ remains below 5.3. For these targets, we also fitted our data with an azimuthally modulated ring that allows non-null closure phases to be produced. This obviously leads to smaller values of $\chi^2_r$ and to similar values of flux contributions and inclinations (Table~\ref{tab:fits_ring_modulated}). Except for VV~CrA~SW for which the position angles are different at a few $\sigma$ level, the position angles agree well between the two models, while the half-flux radii are in agreement within 1-$\sigma$ with those derived from the non-modulated ring.
 
\begin{figure*}[t]
    \centering
    \includegraphics[width=19cm]{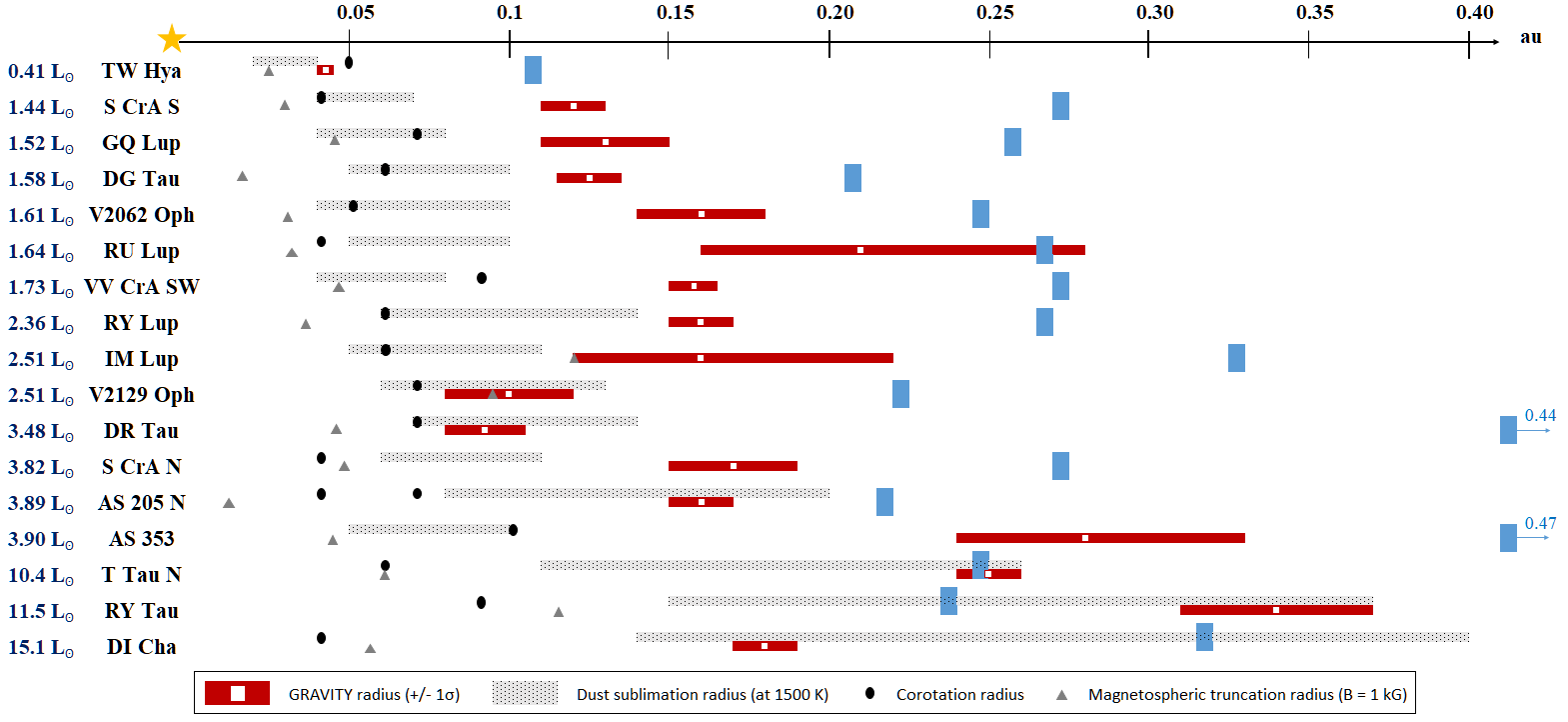}
    \caption{Key size properties of our targets in the ring model approach: corotation radii $R_{\rm co}$ (black circles), magnetic truncation radii $R_{\rm mag}$ for a magnetic field of 1~kG (gray triangles), range of sublimation radii {$R_{\rm sub}$} (gray lines), and half-flux radii $a$ derived from the GRAVITY measurements (white squares) including 1~$\sigma$ error bars (red lines). The width-to-radius ratio $w$ is not included in this schematic view. 
    The targets are ordered by increasing stellar luminosities, from top to bottom. For comparison, we display the achieved angular resolution (blue rectangles) defined by $\lambda_0/2B_{\rm max}$ with $\lambda_0$~=~2.15~$\mu$m the central wavelength of the K band, and $B_{\rm max}$ the maximal interferometric baseline for each data set. All angular resolutions are converted in au by using the distances of Table 1. For DR~Tau and AS~353, the blue arrows indicate the exact location of the blue mark. Two values of $R_{\rm co}$ are reported for AS205\,N from the literature (see Table~\ref{tab:charac_sizes}).}
    \label{fig:CharacteristicSizes}
\end{figure*}

{With the aim to pursue a statistical analysis of our sample, we adopted for all our sources the centro-symmetric geometrical ring model with a halo and considered the values of Table~\ref{tab:fits_ring} to reveal trends within the population of T\,Tauri's disks. }

\subsection{The inner disk morphology}
\label{sect:charac_sizes}

\begin{figure}[h]
    \centering
    \includegraphics[width=7.5cm]{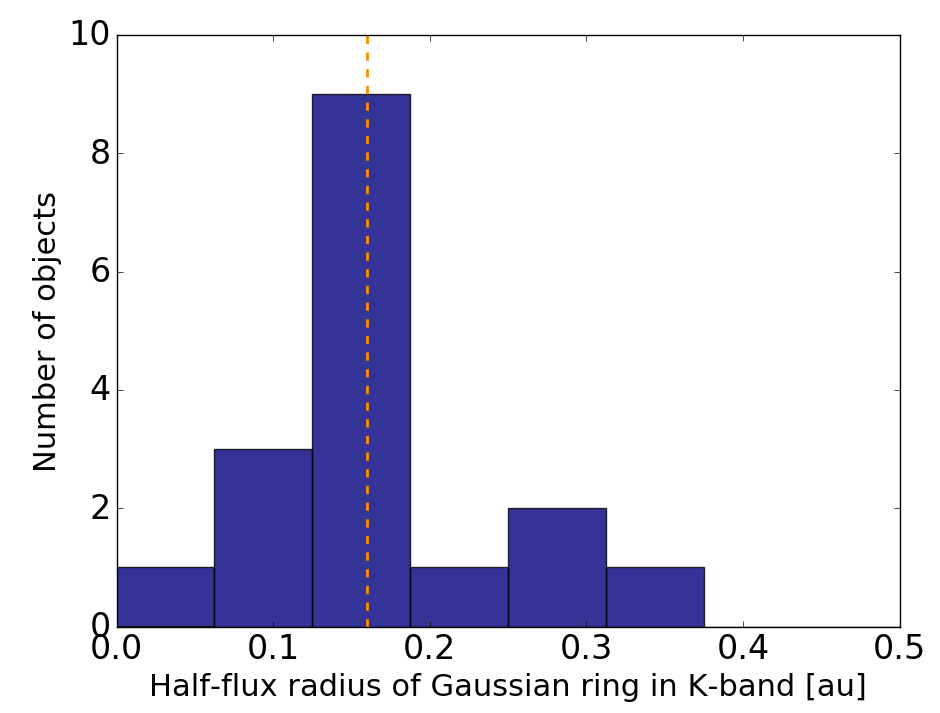}
    \caption{Histogram of the half-flux radius of the ring model in the K band of our sample. The orange dashed line corresponds to the median value.}
    \label{fig:RinK}
\end{figure}

The half-flux radii $a$ of our sample are presented in Fig.~\ref{fig:RinK}. They all remain below 0.34~au, with a median value equal to 0.16~au. This is substantially more compact than what was observed in our Herbig sample spanning a luminosity range of 1--10$^4 L_\odot$, where the K-band half-flux radii range between 0.1 and 6~au, with a median value of 0.6~au. Our T Tauri sample exhibits a median inclination of 34$^\circ$ with values ranging between 14$^\circ$ and 60$^\circ$.
Concerning the width-to-radius ratio $w$, for most of our targets, the error bars are large, and the values for our best fits are consistent with 1 (i.e., with wide rings; Table 2). This is consistent with our angular resolution that limits our capabilities of detecting an inner cavity (i.e., a dust-free region inside the inner front, if any). On {Fig.~\ref{fig:CharacteristicSizes}}, when comparing the half-flux radii (white squares) with the reached angular resolution (blue rectangles), we clearly see that only a few disks are fully resolved (T~Tau~N, RY~Tau, and RU~Lup), as suspected when inspecting the visibility curves (Sect. 3). When our angular resolution is high enough, we are able to disentangle between a smooth rim ($w \sim$~1; RY~Tau) and a sharp rim ($w \sim$~0.1; T~Tau~N).

We compare the half-flux radii derived in the K-band continuum with GRAVITY with the dust sublimation radius $R_{\rm sub}$, inside of which there is no dust, with the corotation radius $R_{\rm co}$ where the Keplerian velocity of the disk equals the stellar velocity, and with the magnetic truncation radius $R_{\rm mag}$ where the stellar magnetic field disrupts the disk.

To determine $R_{\rm sub}$ (in au) for each target of our sample, we used the relation from \cite{Monnier2002}:
\begin{equation}
    R_{\rm sub} = 1.1 \sqrt{Q_R} \sqrt{\frac{L_*}{1000 L_\odot}} \left( \frac{1500}{T_{\rm sub}} \right)^2,
\end{equation}
with $Q_R$ the ratio of the absorption efficiencies of the dust of the incident field and of the reemitted field, and $T_{\rm sub}$ the dust sublimation temperature. 

As emphasized in \cite{Monnier2002}, $Q_R$ depends on the dust properties and on the central star effective temperature (see their Fig. 2). For grain radii ranging from 0.03~$\mu$m to 1~$\mu$m, $Q_R$ ranges from 1 to 4 for stars with $T_{\rm eff}$~=~4000~K; $Q_R$ equals 6 for $T_{\rm eff}$~=~5200~K, and 8 for $T_{\rm eff}$~=~5770~K. We used these values to compute the range of $R_{\rm sub}$ for all our targets when considering $T_{\rm sub}$~=~1500~K. These ranges can only be used as characteristic sizes to which our half-flux radii can be compared since $R_{\rm sub}$ strongly depends on the processes that are taken into account as coupling of grains to gas and/or backwarming by circumstellar dust \citep{Kama2009}. \\


We used the rotational periods $P_{\rm rot}$ provided in the literature (Table~\ref{tab:charac_sizes}) to compute $R_{\rm co}$ through the formula:
\begin{equation}
    R_{\rm co} = (G M_*)^{1/3} (P_{\rm rot}/2\pi)^{2/3}.
\end{equation}
For targets for which the rotational period is not available, we use the $v \sin i$, when available, and consider the inclinations derived from our GRAVITY measurements.

To compute $R_{\rm mag}$ (in $R_\odot$) we used the equation provided by \citet{Hartmann2016}:
\begin{equation}
    R_{\rm mag} = 12.6 \frac{B^{4/7}R^{12/7}}{M^{1/7}\dot{M}_{\rm acc}^{2/7}}.
\end{equation}
We derived the stellar radii $R$ from the luminosities and effective temperatures of Table 1 and we used the masses $M$ and the accretion rates $\dot{M}_{\rm acc}$ of Table 1. As the magnetic field strength $B$ is not known for all our targets, we adopted a fiducial value of  1~kG \citep{Hartmann2016} for all of them to compute $R_{\rm mag}$. 

All the characteristic sizes are reported in Table~\ref{tab:charac_sizes} and displayed in Fig.~\ref{fig:CharacteristicSizes}. The sizes derived from our GRAVITY measurements correspond to half-flux radii of the K-band continuum emission. They are shown in Fig.~\ref{fig:CharacteristicSizes} in the form of white squares, with the extension of the red bars corresponding to the 1~$\sigma$ error bars. These characteristic sizes are, by definition, larger than the sublimation radii. As expected, the stars of our sample with the lowest luminosities (i.e., $\leq$~2~$L_{\odot}$ on the top of the Figure) exhibit the range of sublimation radii (gray lines) closest to the central star, which makes them difficult to resolve, given our angular resolution (blue rectangles). The corotation radii (black circles) and the magnetic truncation radii for 1~kG (gray triangles) are generally of the same order of magnitude (e.g., a few stellar radii), and much smaller than the half-flux radii of the K-band continuum emission measured with GRAVITY.

\subsection{Comparison with previous measurements by near-infrared interferometry}


Seven stars of our sample are part of the PIONIER survey led by \cite{Anthonioz2015}. The authors fit their data with a model made of three components: the central star, a thermal emission ring whose brightness distribution is constant and whose width-to-radius ratio is fixed to 0.18, and a scattered light ring (see Sect. 6.3). Both rings have the same inner radius. The ring inclination is fixed to 0$^\circ$, which is not a realistic assumption for our targets (see Sect. 5.2). When comparing the PIONIER inner rim radii corrected from the Gaia EDR3 distances with the GRAVITY half-flux radii (Fig.~\ref{fig:PIONIER}), the H-band sizes appear smaller than the K-band ones for five of the seven objects, as expected since the PIONIER model is based on a ring with an inner cavity without emission inside, while the GRAVITY fits provide half-flux radii. AS~205~N exhibits H-band and K-band sizes that only differ at a 2.6-$\sigma$ level but the fit of the PIONIER data has a very large $\chi^2$ (see Table~\ref{tab:PIONIER}). The disk of TW~Hya is almost unresolved with PIONIER whose visibilities squared are consistent with 1, even for the longest baselines, while the visibilities squared of GRAVITY go down to 0.8. This is in agreement with a smaller flux contribution of the ring (whose size is unconstrained) in the H band (3\%) than in the K band (14\%). 

 
Six targets of our sample have been previously observed in the K band with the two telescopes of the KI by \citet{Eisner2010,Eisner2014}. In both papers, the authors report the K-band inner radii of a ring model, whose width-to-radius ratio equals 0.2, and do not take into account the potential scattered light contribution. For some objects they report different values between the two papers. For DG~Tau, a size measurement has also been published by \cite{Akeson2005}, when considering a model with the central star, an incoherent contribution from scattered light, and a resolved ring. All the angular diameters derived from the KI and the corresponding radii in au, when corrected for the Gaia EDR3 distances, are given in Table~\ref{tab:KI}. When comparing the GRAVITY half-flux radii with the measurements by \cite{Eisner2010,Eisner2014}, three stars have comparable sizes, and two have a half-flux radii about twice as large as the inner radii of the annular ring (Fig.~\ref{fig:KI}). DR~Tau appears as an outlier, with an inner radius of the annular ring about twice smaller than the half-flux radius derived from GRAVITY, while the flux contributions for the disk are almost the same (about 80\% for KI, and 90\% for GRAVITY). As the visibilities obtained with the KI on this star are not plotted in the papers, it is impossible to compare their values with ours. However, from their log, we notice that their two-telescope observations only span the 37~M$\lambda$ spatial frequency and a 15$^\circ$ range of baseline orientations, which could explain why the inclination the authors derive (45$^\circ$) is quite different from the GRAVITY one (about 20$^\circ$). It is worth mentioning that the oscillations observed in the GRAVITY visibilities also point toward a complex environment, for which our geometrical model might be too simplistic ($\chi^2$~$\sim$~3). This is not surprising since DR~Tau is known to be an active T Tauri star that exhibits strong photometric and spectroscopic variations, suggesting very dynamic interactions between the star and its disk \citep{Alencar2001,Banzatti2014}. For DG~Tau, the GRAVITY determination and the KI one by \cite{Akeson2005} match perfectly (see Sect. 6.3 for the discussion about the scattered light contribution).




\begin{figure*}[t]
    \centering
    \includegraphics[width=13.7cm]{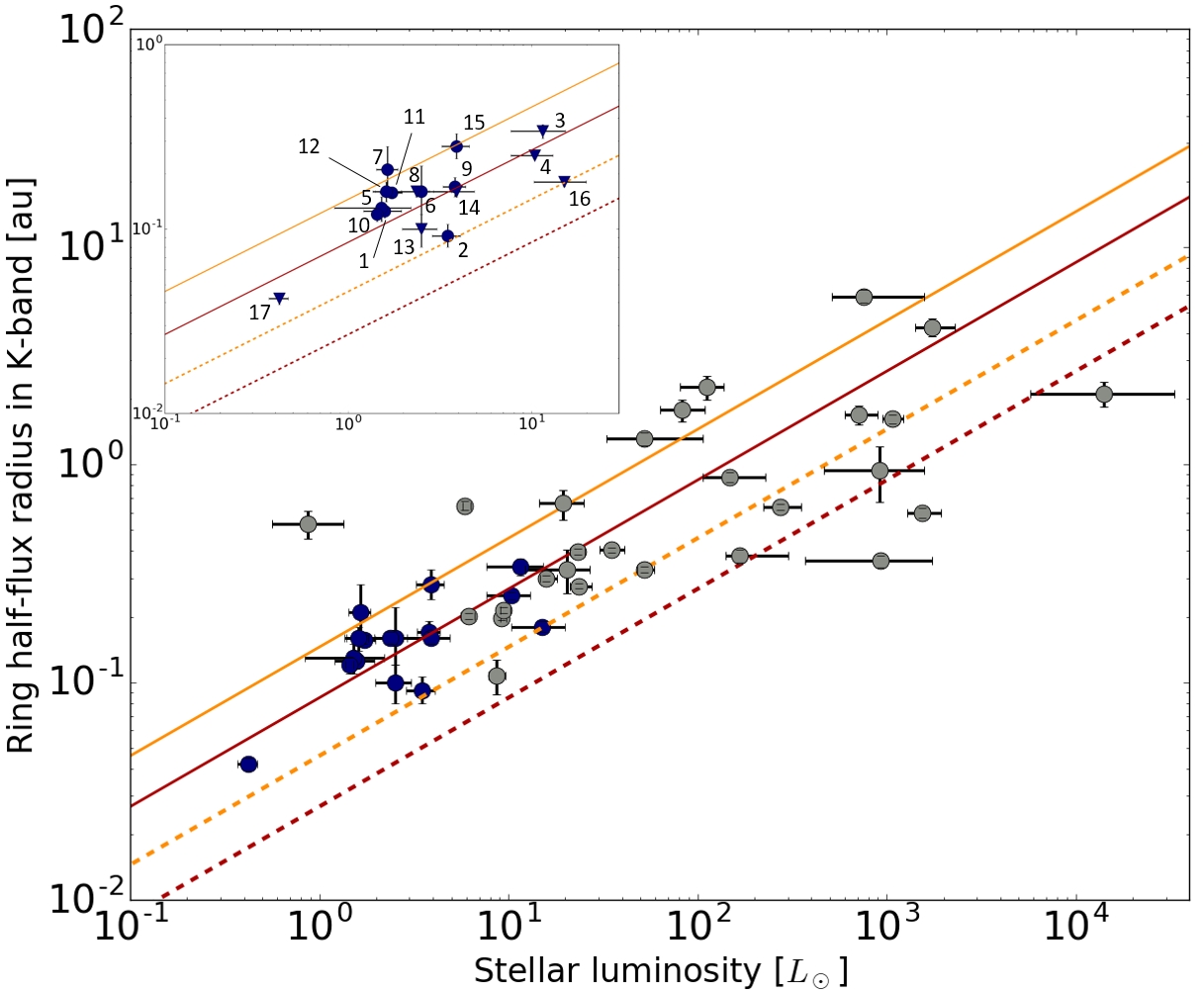}
    \caption{Radius-luminosity relation for our T Tauri sample (blue symbols) and for the Herbig sample (gray symbols; see GC19). The lines correspond to different models of a passively irradiated disk with an optically thin inner cavity: dust grain temperature of 1300~K (orange) and of 1700~K (red); dust cooling efficiency of 0.1 (solid lines) and of 1 (dashed lines). The insert zooms in on the T Tauri region: the numbers refer to Table 1; the circles denote the fully convective stars, while the triangles denote the stars with a radiative core and a convective envelope.}
    \label{fig:LRrelation}
\end{figure*}

\section{Discussion}

\subsection{The Radius-luminosity relation}

We gather all our measurements from this study and GC19 in a size-luminosity diagram (Fig.~\ref{fig:LRrelation}) that highlights the correlation between the location of the K-band emission and the luminosity of the central star. We indicate the predictions for a passively irradiated disk with an optical thin inner cavity for two different dust grain temperatures $T_g$ (1300~K and 1700~K, in orange and red, respectively), and two extreme dust cooling efficiencies $\epsilon$ (0.1 and 1, in solid and dashed lines, respectively); the radius scales as $L^{1/2}\epsilon^{-1/2}T_g^{-2}$ when assuming a backwarming coef\-ficient equal to 1. If the inner dusty disk is truncated by dust sublimation close to the star, the higher the stellar luminosity, the larger the half-flux radii in the K band. 

GRAVITY measurements allow us to extend the size-luminosity relation (hereafter the R-L relation) toward the lower luminosity limit with respect to GC19. While the GRAVITY measurements follow the general trend $R \propto L^{1/2}$, we notice a large scatter at high luminosities (around 10$^3 L_\odot$) that was already pointed out and discussed in GC19, and a hint of a departure from the R-L relation for luminosities below 1~$L_\odot$. Most targets with luminosities in the range of 1-10~$L_\odot$ exhibit similar half-flux radii in the K band around 0.1-0.15~au, and for a given stellar luminosity, the half-flux radii span a large range of sizes. Such a trend for T Tauri stars was mentioned already in \cite{Akeson2005} and \cite{MaccAS205}, and several phenomena have been invoked to explain this departure. They include peculiar dust properties (lower sublimation temperatures, smaller and thus hotter grains), extra heating induced by the viscous energy dissipation from accretion, and inner disk position controlled by the stellar magnetosphere pressure \citep{Eisner2007}. \cite{Pinte2008} instead suggest that the position of the inner disk as determined by long-baseline interferometric measurements might be incorrectly estimated, especially for T Tauri stars, when the contribution of scattered light is neglected. We investigate these assertions in the following sections.

\subsection{Impact of the magnetic field and accretion rate}

In their sample of T Tauri stars, \cite{Eisner2007} observe a larger departure from the R-L relation for those targets that exhibit lower accretion luminosities (see their Fig. 6). These authors suggest that the mechanism that leads to the truncation of the dusty inner disk may depend on the accretion rate. They advocate that magnetic truncation can explain the large inner disk radii of low-mass T Tauri stars, as low accretion rates can cause magnetospheric truncation outside of the dust sublimation radius \citep{Eisner2006}. Such a question is difficult to address when the targets are only partially resolved. This is our case and the case of Eisner's observations too (their squared visibilities are generally larger than 0.8), in particular for the low-luminosity stars whose inner rims might be very close to the central stars and appear very compact. Nevertheless, to test this hypothesis, we gather the mass accretion rates $\dot{M}_{\rm acc}$ from the literature (Table~\ref{tab:param_fond}). Even if the error bars on $\dot{M}_{\rm acc}$ can be quite large, our sample contains weak accretors with $\dot{M}_{\rm acc}$ lower than 10$^{-8}$~$M_{\odot}$.yr$^{-1}$ (TW~Hya, RY~Tau, RY~Lup, V2129~Oph), moderate accretors with $\dot{M}_{\rm acc}$ of several 10$^{-8}$~$M_{\odot}$.yr$^{-1}$, and strong accretors with $\dot{M}_{\rm acc}$ larger than 10$^{-7}$~$M_{\odot}$.yr$^{-1}$ (DG~Tau, RU~Lup, S~CrA~N, AS~205~N). 

\begin{figure}[h]
    \centering
    \includegraphics[width=8cm]{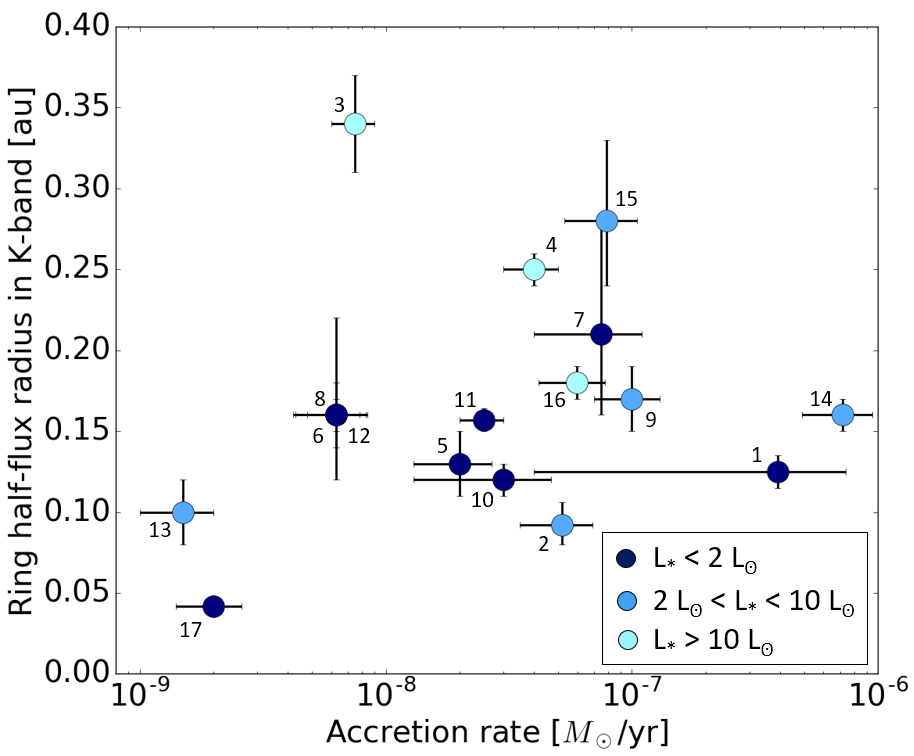}
    \caption{Half-flux radius of the ring model in the K band as a function of the accretion rate of our targets. The color codes the luminosity of the central star and the numbers refer to Table~\ref{tab:param_fond}.}
    \label{fig:RinKvsMacc}
\end{figure}

\begin{figure*}[t]
    \centering
    \includegraphics[width=9.1cm]{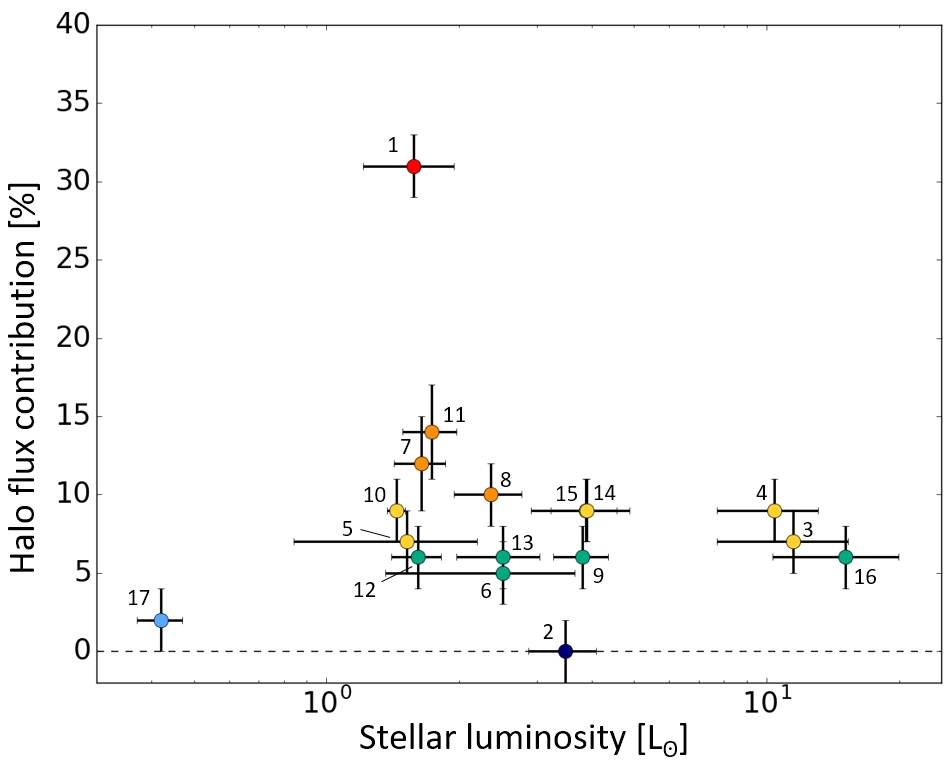}
    \includegraphics[width=9cm]{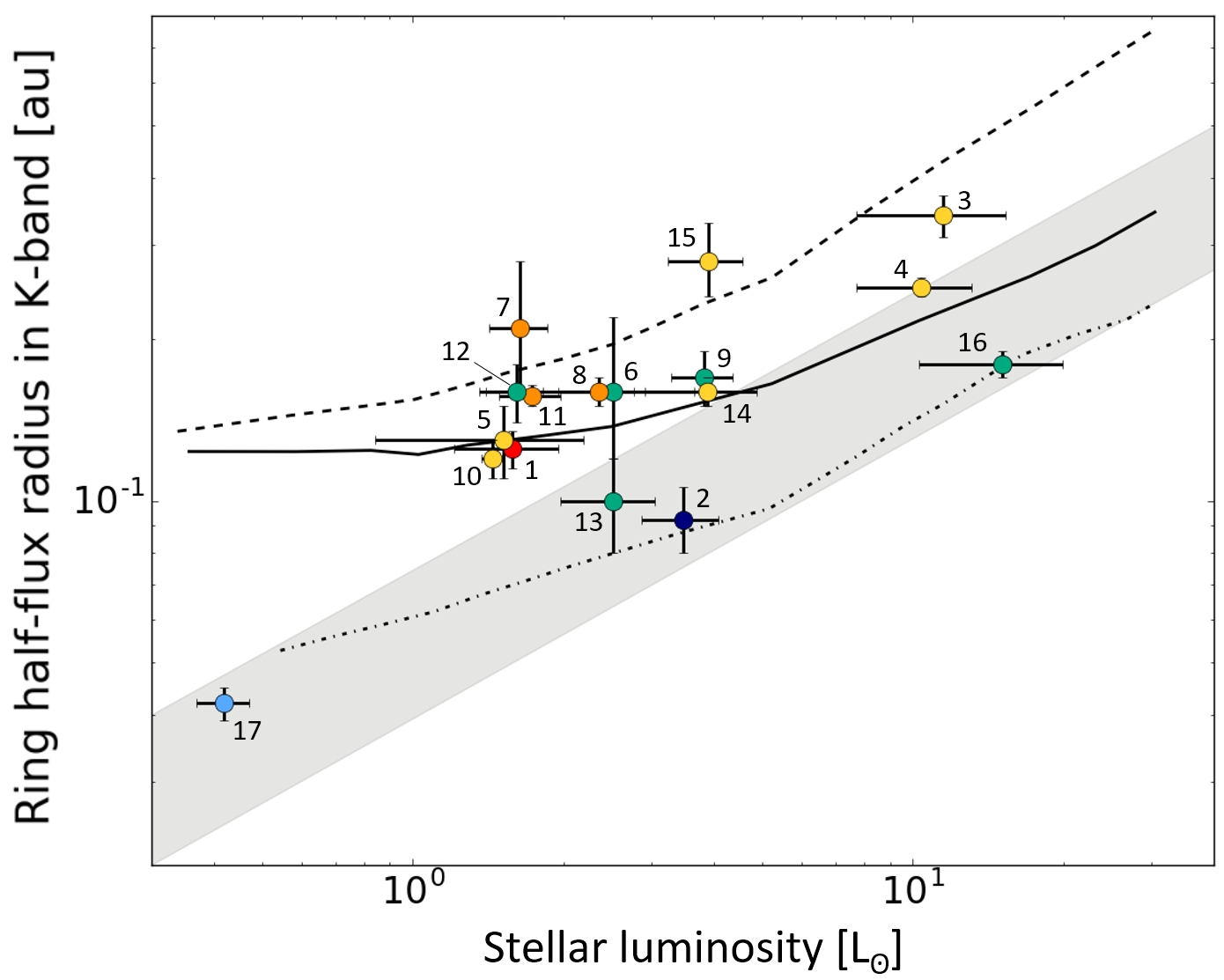}
    \caption{Halo contribution (left) and half-flux radius (right) of the best-fit ring model in the K band as a function of the luminosity of the central star. The color codes the amount of halo contribution and the numbers refer to Table~\ref{tab:fits_ring}. On the right plot, the sublimation radius is computed for a grain sublimation temperature between 1500~K and 2000~K (shaded area). Different grain population models as computed in \cite{Pinte2008} are reported: Model A (amorphous silicate grains smaller than 1~$\mu$m) at 1500~K (dashed line) and 2000~K (solid line), and Model D (amorphous silicate and carbon grains smaller than 1~mm; dash-dotted line). All of these models include the scattered light processing.}
    \label{fig:halo}
\end{figure*}

The GRAVITY half-flux radii as a function of the accretion rate appear to be almost constant for the different subclasses of stellar luminosity, even when considering only the more luminous and thus the most resolved stars (in light and medium blue in Fig.~\ref{fig:RinKvsMacc}). When plotting the GRAVITY half-radius as a function of the ratio between the accretion luminosity and the stellar luminosity, we do not detect a trend either. We do not see the correlation pointed out by \cite{Eisner2007}, who suggest that at sufficiently low accretion rates, magnetic truncation radii $R_{\rm mag}$ can become larger than $R_{\rm sub}$, leading to a larger K-band size \citep{MaccAS205}. As mentioned in Sect.~\ref{sect:charac_sizes}, the magnetic truncation radii $R_{\rm mag}$ are generally much smaller than the half-flux radii derived with GRAVITY. Since it scales as $B^{4/7}$, having $R_{\rm mag}$ as large as the half-flux radii measured with GRAVITY would imply very strong magnetic fields (i.e., larger than several kG) that have not been detected so far. Finally, when looking for a trend with the magnetic properties of the central star (see insert of Fig.~\ref{fig:LRrelation}), no clear difference emerges between the fully convective stars (circles) and those with a radiative core and a convective envelope (triangles), in which the dipolar component of the magnetic field is strong (on the order of 1~kG). At our spatial resolution, the magnetic and accretion properties do not appear to be the main parameters that govern the location of the half-flux radius of the inner dusty disk.



\subsection{Contribution of the scattered light}

Radiative transfer simulations done with MCFOST by \cite{Pinte2008} show that, for the T Tauri stars in the K band, the scattered starlight and the thermal emission are of the same order of magnitude. The reasons are twofold: these cooler, less massive stars radiate a larger fraction of their luminosity in the near-infrared range, and the K-band emitting region of the disk is more compact (cooler disk with a smaller inner radius). As emphasized by the authors, since the direct thermal emission decreases radially faster than the scattered light, this latter component can play a significant role in the visibilities. Indeed, scattered light is spatially extended and could be much more extended than the angular resolution of the interferometer, and thus fully resolved by all the interferometric baselines. This implies a steep decrease in visibility for very short baselines due to its flux contribution (see Fig. 2 of \cite{Pinte2008}). This steep decline has been already observed in the H band with PIONIER by \cite{Anthonioz2015} for three T Tauri stars for which observations at short baselines (10-20~m) are available (their Fig. 2). Despite the lack of GRAVITY observations at such short baselines, the visibility curves (Figs. B.1-B-4) clearly show that, for some objects, the visibilities do not reach 1, when extrapolated toward the shortest baselines. The most obvious cases are DG~Tau, RU~Lup, RY~Lup, and VV~CrA~SW. \cite{Pinte2008} pointed out that the K-band radius of the inner disk derived from a model that only includes a Gaussian ring is always overestimated, and that this overestimation depends on the field-of-view of the telescopes and increases as the temperature of the central object decreases (see their Fig. 3). 

While developing full radiative transfer models of all the targets is beyond the scope of this paper, we take into account the scattered light contribution by including an extended contribution, namely the halo in our model (see Sect.~\ref{sect:model}). Due to its fractional flux contribution (i.e., $f_s$~+~$f_c$~+~$f_h$~=~1), its net effect is that for very short baselines, the visibility is smaller than 1. When fitting our ring model, the halo flux contribution for our sample has a mean of 8.7\% and a median of 7\% (see column 6 of Table~\ref{tab:fits_ring}). The strongest halo contributions (about 10\% or more) are observed for the four targets whose visibility curves exhibit visibilities different from 1 at the shortest baselines: DG~Tau whose halo contribution reaches about 30\% (\#1 in Fig.~\ref{fig:halo}-left), RU~Lup (\#7), RY~Lup (\#8), and VV~CrA~SW (\#11). 
{For these targets, the addition of such a component clearly improves the fit quality (see last column of Table~\ref{tab:fits_ring}). These findings reinforce the idea, as suggested by previous studies as well, that the addition of a fully resolved halo component is often required in the context of near-infrared interferometric studies of YSO disks. The astrophysical interpretation of the halo component might be, however, subtler.}

Probably due to the star-to-star differences (such as different grain sizes and/or materials) and the limited number of our targets, there is no clear trend of this halo flux contribution with the luminosity of the central star (Fig.~\ref{fig:halo}-left). 
Fig.~\ref{fig:halo}-right displays the half-flux radii measured with GRAVITY as a function of the stellar luminosity. Since we accurately determine the orientations of the inner disks, the observed scatter cannot be explained by neglected inclination effects. At low luminosity (i.e., around 1-2~$L_\odot$ or less), only TW~Hya (\#17), which exhibits a very small halo flux contribution (2~$\pm$~2\%), is in agreement with the dust sublimation radius computation (shaded area). For the others, whose halo flux contributions span between 6 to 31\%, the measured sizes are larger than the dust sublimation radius. The departure is in good agreement with the different models computed by \cite{Pinte2008} for different dust populations. These results nicely support the claim by \cite{Pinte2008} and what has been previously validated for several T Tauri stars in the H band with PIONIER by \cite{Anthonioz2015}, even if all the sizes were determined for a face-on inclination. 
 Despite its simplicity, the halo model might be seen as a proxy for the importance of the scattered light in these disks. We see that, in particular for RU~Lup (\#7), which is well resolved, accounting for the halo flux contribution ($f_h$~=~12\%) is insufficient to bring the object down in the gray shaded area of the R-L relation, contrary to what we may have expected. This could indicate that the halo model is not accurate enough in describing the scattered light contribution and/or that the object harbors a larger inner disk due to a specific mechanism (e.g., clearing or viscous heating by accretion). {DR~Tau (\#2) is another interesting case since a null halo contribution is detected. Its position on  Fig.~\ref{fig:halo}-right suggests that the dust around this object might be amorphous silicate and carbon grains with a maximal size of 1~mm (Model D of Pinte et al. (2008) displayed by the dash-dotted line) whose scattering properties (albedo and phase functions) are different from those of the silicate grains. It is clear that a few objects in our sample deserve a more sophisticated radiative transfer model -- as performed on RY~Tau by \citet{Davies2020}, and on RY~Lup by \citet{Bouarour2020} -- to shed more light on these objects.}

\subsection{Inner disk morphology and evolution}

Mid-infrared interferometry allows us to trace dust at temperatures down to $\sim$~300~K and probe the temperature profile in the disk. This wavelength regime also probes the disk regions on a few au scale, beyond the dust sublimation rim, and thus can be used to investigate the flaring of the disks. The N band also includes the silicate feature and thus allows the silicate dust to be probed. The shape of this feature is related to the dust composition and the grain size distribution \citep{Natta2007,Henning2010} and can provide information about dust evolution as coagulation and crystallization \citep{vanBoekel2003,VB2005A&A...437..189V}. Finally, the absence and/or weakness of the N-band silicate feature is associated with the presence of large gaps in the radial dust distribution or of large dust-depleted inner regions of protoplanetary disks \citep{Maaskant2013A&A...555A..64M}. Several scenarios on the potential evolutionary sequence between the different disk morphologies are still under debate: for the Herbig stars, between the ungapped/gapped flat disks and flared disks \citep{Maaskant2013A&A...555A..64M,Menu2015A&A...581A.107M}; for T Tauri stars, between the primordial disks and the depleted disks or the disks with large inner holes or gaps \citep{Currie2009}.

\begin{table}[t]
\centering
\caption{K-band and N-band characteristic sizes. Near-infrared sizes correspond to half-flux radius of a ring model. Mid-infrared data are the half-flux radius adapted from \cite{Varga2018}. Multiplicative factor $f$ is applied to the measured N-band cha\-racteristic size. The first column reports the numbering of Table 1. The last column gives the type of the silicate feature (a: absorption; e: emission; w: weak or flat) as provided in Table E.1 of \cite{Varga2018}.}
\begin{tabular}{c c c c c c}
\hline
 \# & Object & $HWHM_K$ & $HWHM_N$ & $f$ & Silicate \\
  &      & [mas]     & [mas]     &   & Feature\\ \hline
 1 & DG~Tau  & 1.0~$\pm$~0.1  & 8.0$\pm$0.6   & 1.54 & wa \\ [1ex]
 2 & DR~Tau  & 0.48$^{+0.07}_{-0.06}$  & 2.1$\pm$0.4  & 1.05 & e \\[1ex]
 3 & RY~Tau   & 2.45$^{+0.24}_{-0.22}$ & 8.0$\pm$0.7 & 1.50 & e \\[1ex]
 4 & T~Tau~N   & 1.74$\pm$0.08 & 6.3$\pm$0.3    & 1.18 & e \\[1ex]
 7 & RU~Lup  & 1.32$^{+0.42}_{-0.32}$  & 4.7$^{+0.6}_{-0.1}$ & 1.04 & e\\[1ex]
 9 & S~CrA~N   & 1.05$^{+0.15}_{-0.14}$ & 10.2$\pm$1.7 & 1.11 & ae \\[1ex]
 10 & S~CrA~S   & 0.83$^{+0.08}_{-0.07}$ & 4.4$\pm$0.6 & 1.16& ae\\[1ex]
 11 & VV~CrA~SW   & 1.00$\pm$0.05 & 7.4$^{+1.2}_{-0.6}$  & 1.15 & a \\[1ex]
 12 & V2062~Oph & 1.07$^{+0.13}_{-0.12}$& 3.9$\pm$0.6$^{*}$ & 1.15 & e$^{+}$ \\[1ex]
 13 & V2129~Oph  & 0.79$^{+0.14}_{-0.13}$ & 12.4$\pm$3.6 & 1.33 & e \\[1ex]
 14 & AS~205~N  & 1.17$^{+0.09}_{-0.07}$ & 6.9$^{+1.7}_{-0.9}$ & 1.39 & e\\[1ex]
 16 & DI~Cha  & 0.95$\pm$0.05 & 13.9$^{+1.6}_{-1.1}$ & 1.06 & wa\\[1ex] 
 17 & TW~Hya  & 0.69$\pm$0.05 & 29.4$\pm$1.7  & 1.03 & w \\[1ex]
\hline
\end{tabular}\label{tab:NtoK}

\tablefoot{$^{*}$ from \cite{Menu2015A&A...581A.107M}; $^{+}$ from \cite{Furlan2009}.}
\end{table}

\begin{figure}[b]
    \centering
    \includegraphics[width=7.5cm]{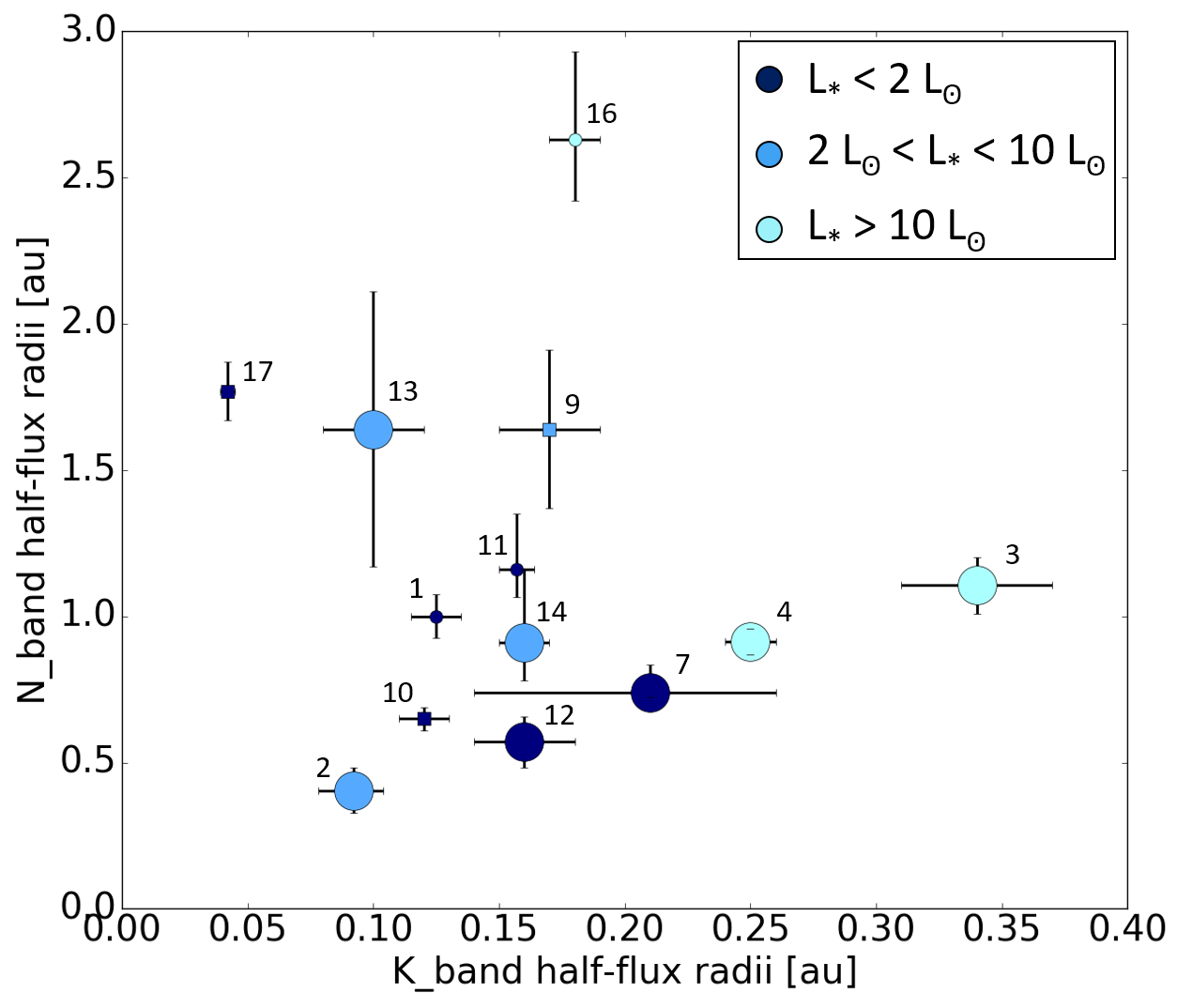}
    \caption{N-band size as derived from MIDI measurements \citep{Varga2018,Menu2015A&A...581A.107M} as a function of K-band size derived with GRAVITY for the T Tauri sample. The color codes the luminosity of the central star and the numbers refer to Table~\ref{tab:NtoK}. The circles correspond to the stars exhibiting a N-band silicate feature in emission (large circles) or in absorption (small circles). The squares correspond to the stars for which the silicate feature is variable or weak.}
    \label{fig:NandKsizes}
\end{figure}
We used the mid-infrared half-flux radii derived from MIDI observations and reported in \cite{Varga2018} to compare them with our half-flux radii in the K band for 12 of our targets (Table~\ref{tab:NtoK}). We also added the N-band size for V2062~Oph derived  by \cite{Menu2015A&A...581A.107M}. Since these authors model the targets as face-on disks, we corrected for the inclination effect\footnote{As in GC19, we correct with a multiplicative factor $f$: for a uniform {\it (u, v)} coverage, $f$~=~$2 - \cos i$, with $\cos i$ determined from our GRAVITY measurements; for a {\it (u, v)} plane with a preferential baseline along the ellipse minor axis $f$~=~$1/\cos i$.}. From Table~\ref{tab:NtoK} and the Gaia EDR3 distances, we computed the N-band and K-band sizes in au. As a potential proxy of the disk morphology, we also gather the type of the silicate feature as determined by \cite{Varga2018} from the analysis of the MIDI spectra (last column in Table~\ref{tab:NtoK}). For our sample, Figure~\ref{fig:NandKsizes} suggests that the stars with a silicate feature in emission (large circles) have N-band sizes about 3 times larger than the K-band ones. V2129~Oph (\#13) appears as an outlier, but for this target the MIDI data set contains only one observation, leading to the largest error bar on the N-band size. For stars with silicate features in absorption (small circles) that could be associated with disks with large gaps, the scatter in the N-band size versus K-band size plot is larger. At small scales ($<$~0.16~au in the K band and $<$~1.2~au in the N band), both populations (i.e., with silicate feature in emission and in absorption) are present. These findings are quite similar to what was observed for the Herbig stars with gapped and flat disks (GC19). 
No clear correlation with the stellar luminosity is observed (Fig.~\ref{fig:NandKsizes}), which is not surprising since, for the T Tauri stars, there is a clear departure from the R-L relation in the near-infrared (Sect. 6.1) and in the mid-infrared ranges \citep{Menu2015A&A...581A.107M, Varga2018}.

\begin{table*}[t]
\centering
\caption{Inclinations and position angles of the inner disks as derived from GRAVITY observations (this work) and of the outer disks as derived from ALMA observations for nine targets of our sample. The first column reports the numbering of Table 1. The last column gives the corresponding references for the ALMA estimations.}
\begin{tabular}{ccccccc}
\hline
 \# & Object & $i_{\rm in}$ [$^\circ$] & $PA_{\rm in}$ [$^\circ$] & $i_{\rm out}$ [$^\circ$] & $PA_{\rm out}$ [$^\circ$] & References \\
\hline
 1 & DG~Tau  & 49~$\pm$~4  & 143$\pm$12 & 41~$\pm$~2  & 128$\pm$16   & \cite{Podio2019} \\
 3 & RY~Tau  & 60~$\pm$~1 & 8~$\pm$~1 & 65 & 23 & \cite{Francis2020}   \\
 5 & GQ~Lup  & 22~$\pm$~6 & 180~$\pm$~3 & 60.5~$\pm$~0.5 & 346~$\pm$~1 & \cite{MacGregor2017} \\
 6 & IM~Lup & 59~$\pm$~4 & 139~$\pm$~3 & 47.5~$\pm$~0.5 & 144~$\pm$~0.7 & \cite{Huang2018} \\ 
 7 &  RU~Lup & 16~$^{+6}_{-8}$  & 99~$\pm$~31 & 18.5~$\pm$~2  & 121.5~$\pm$~7 & \cite{Huang2018}\\
 8 & RY~Lup  & 53~$\pm$~5 & 73~$\pm$~2 & 67 & 109 & \cite{Francis2020} \\
 12 & V2062~Oph & 32~$\pm$~4 & 137~$\pm$~4 & 20 & 30 & \cite{Francis2020}\\
 14 & AS~205~N& 44~$\pm$~2 & 90~$\pm$~1 & 20.1~$\pm$~3.3 & 114.0~$\pm$~11.8 & \cite{Kurtovic2018} \\ 
 17 & TW~Hya & 14~$^{+6}_{-14}$ & 130~$\pm$~32 & 7 & 155  & \cite{Francis2020}  \\
\hline
\end{tabular}
\label{tab:outerdisk}
\end{table*}

\begin{figure}[h]
    \centering
    \includegraphics[width=8.1cm]{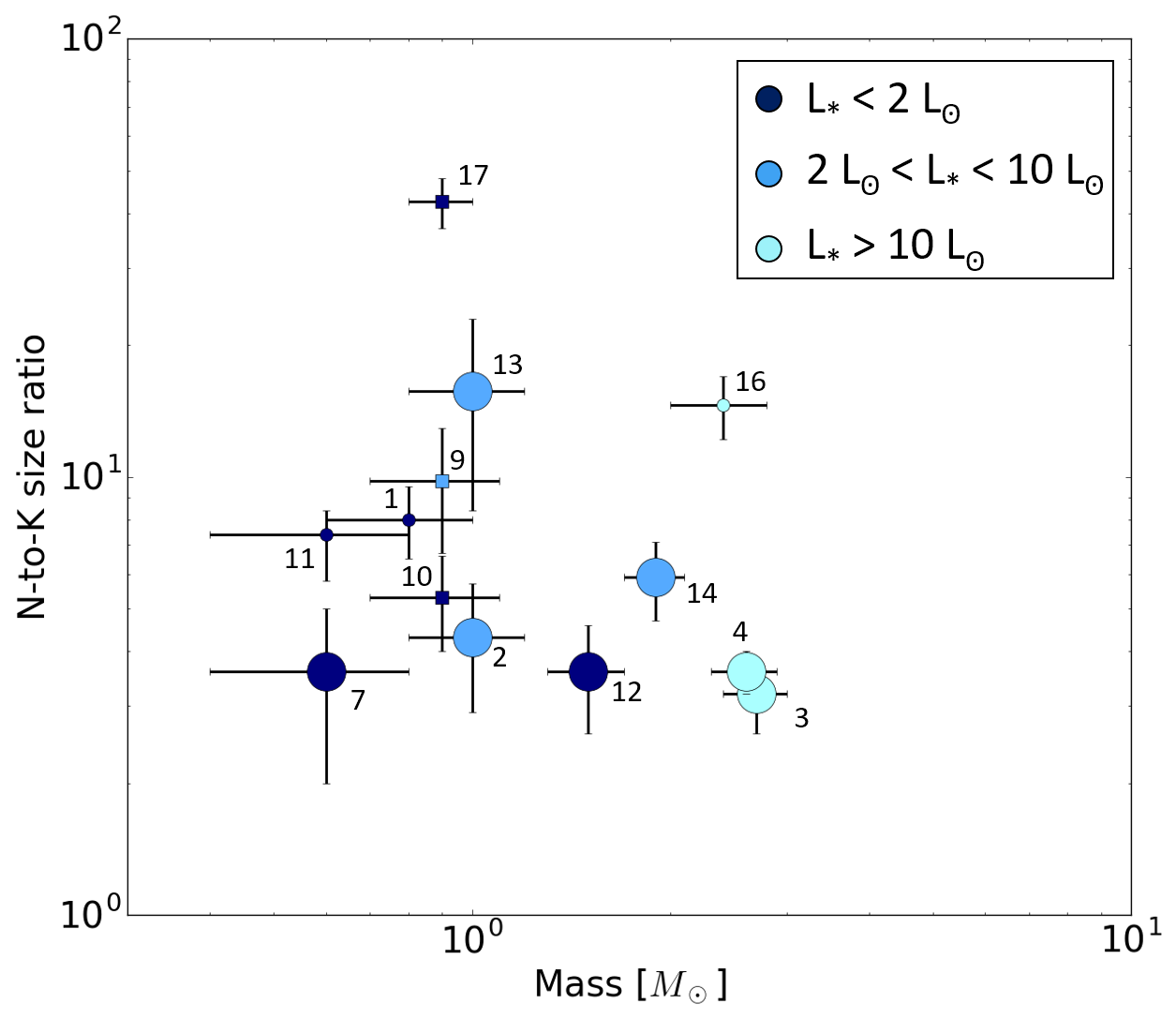}    \includegraphics[width=7.8cm]{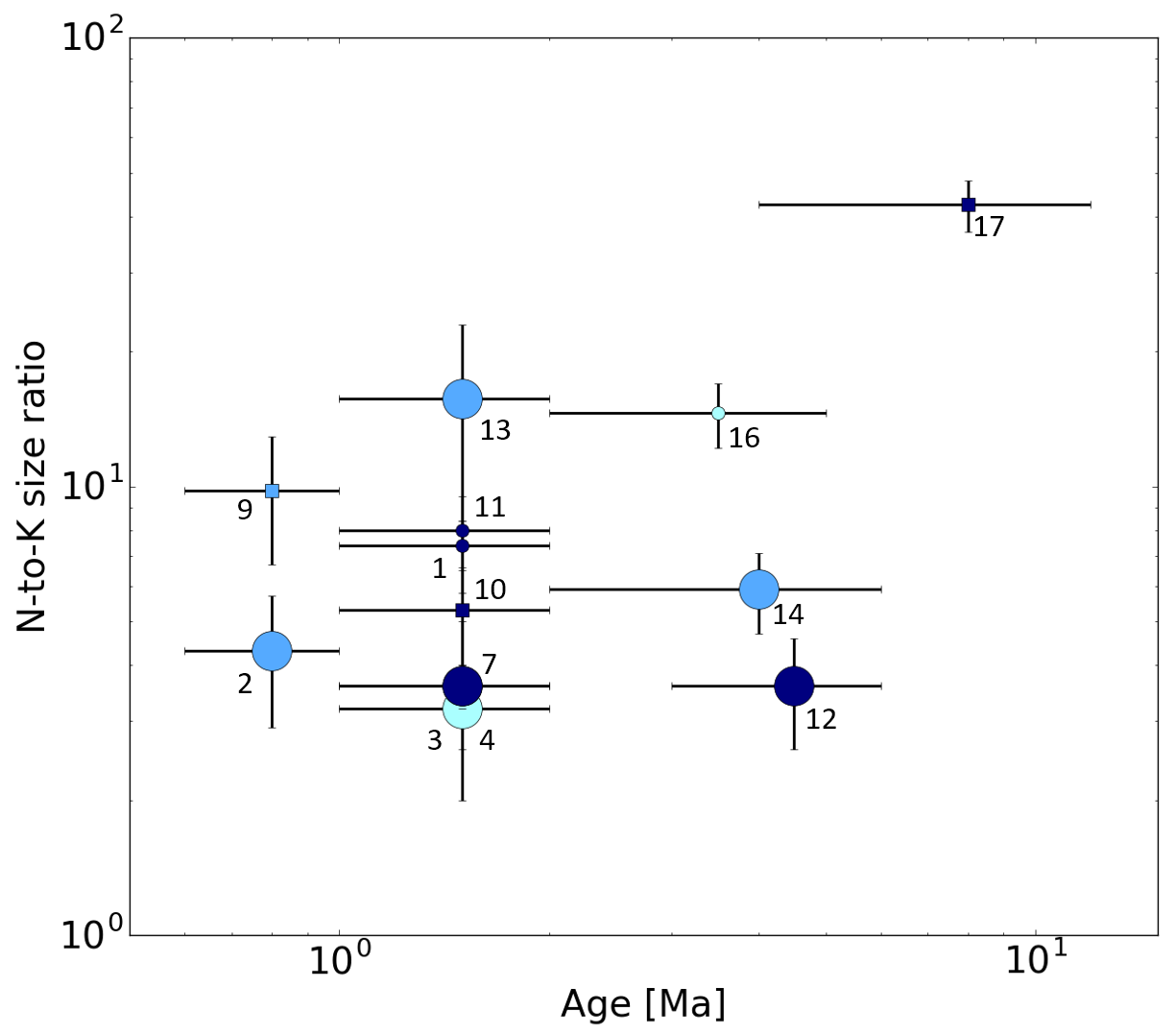}
    \caption{N-to-K size ratio for the T Tauri sample as a function of mass (top) and age (bottom). The color codes the luminosity of the central star and the numbers refer to Table~\ref{tab:NtoK}. The circles correspond to the stars exhibiting a N-band silicate feature in emission (large circles) or in absorption (small circles). The squares correspond to the stars for which the silicate feature is variable or weak.}
    \label{fig:NtoKratio}
\end{figure}

We also computed for each target the size ratio between the N band to the K band and plotted them as a function of age and mass, as we did for the Herbig stars. Since we cannot find a single set of PMS evolutionary tracks to derive mass and age for both the Herbig and the T Tauri stars, we cannot plot all our targets in the same figure because the masses and the ages might be not consistent. We can only check whether the same trends are observed. For Herbig stars, we observe that the N-to-K size ratio remains below 10 for the brightest and most massive Herbig stars, while the objects with lower luminosities (3-80~$L_\odot$) exhibit a stronger scatter in the N-to-K size ratio with values as high as 100. We also observe an increasing scatter in the N-to-K size ratio for stars older than a few Ma and for the gapped disks. We thus suggest {in GC19} that the K-band to N-band size ratio could be used as a proxy to disentangle flat and flared disks. For the T Tauri stars, there is no clear trend with the mass or the luminosity (Fig.~\ref{fig:NtoKratio}-top); for stars of a few Ma, the size ratio remains around 10 or less, while the highest N-to-K size ratio (larger than 40) is observed for the oldest star, TW~Hya (\#17 in Fig.~\ref{fig:NtoKratio}-bottom). As mentioned previously the smallest N-to-K size ratio (around 3) is observed for stars with silicate {emission} features over a mass range of 0.6--2.7~$M_\odot$ and over an age range of 1-6~Ma. 

\subsection{Inner and outer disk (mis)alignment}

In addition to revealing various structures as rings and gaps in the outer parts of protoplanetary disks out to a few hundred au, ALMA images also allow the inclinations and the position angles of the outer disks to be determined \citep[see for instance][]{Long_2018}. We can thus use our GRAVITY measurements to compare the inner and outer disk orientations, and thus better understand the dynamical effects in the disks, and the variety and the complexity of the observed features (like spirals, warps, shadows) by direct imaging and sub-millimetric interferometry \citep{Benisty2015A&A...578L...6B, Benisty2017A&A...597A..42B,Andrews2018Msngr.174...19A,Long_2018,Avenhaus2018ApJ...863...44A}. Among our sample, nine T Tauri stars have been observed with ALMA. We gathered the orientations of their outer disks in the literature (Table~\ref{tab:outerdisk}) and directly compare them with our measurements (third and fourth columns of Table~\ref{tab:fits_ring}). To be conservative, we consider floor error bars of 5$^\circ$ on inclinations and position angles. 

\begin{figure}[h]
    \centering
    \includegraphics[width=9cm]{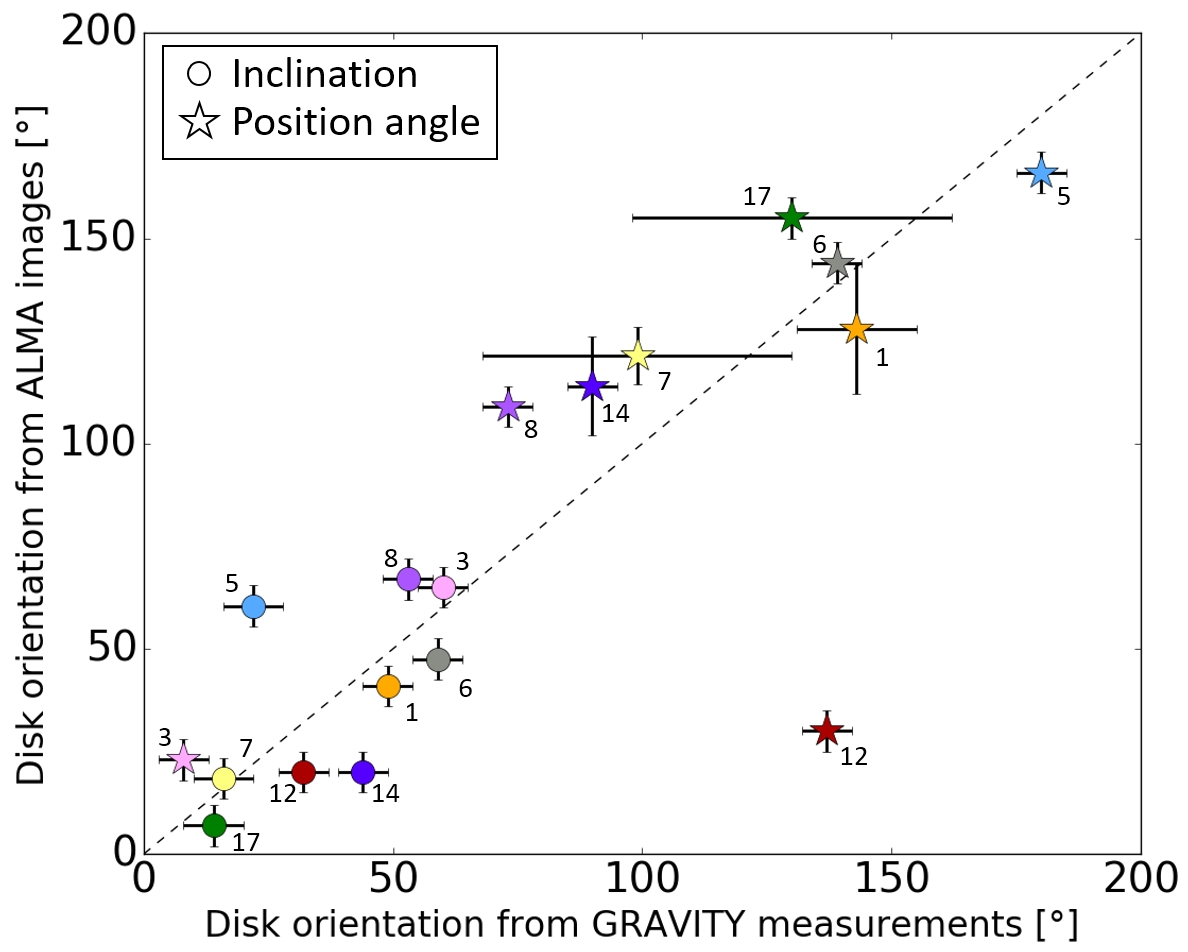}
    \caption{Comparison of the orientations of the outer disk derived from ALMA images as a function of those of the inner disk derived from GRAVITY interferometric measurements. For each object that is marked by a different color, we denote the inclinations with circles, and the position angles with stars. The dashed line shows the 1:1 relation. The numbers refer to Table~\ref{tab:outerdisk}.}
    \label{fig:inner-outer}
\end{figure} 
For five targets, there is a good agreement of the orientations for the inner and the outer disks but the data do not exclude possible small differences in inclination and position angles of about 10-20$^\circ$ (Fig.~\ref{fig:inner-outer}). In particular, for stars almost seen pole-on (i.e., with small inclinations), the position angle is not accurately determined. Clearly there are four outliers: 

\vskip 0.05cm
\textbf{GQ~Lup (\#5)}.
This system is composed of the primary star, GQ~Lup~A, a classical T Tauri star with a circumstellar disk, and a sub-stellar companion, GQ~Lup~b \citep{Neuhauser2005}, whose projected separation from GQ~Lup~A equals 0.7" \citep{Ginski2014}, whose orbit might be highly eccentric and whose semi-major axes might range 100-185~au \citep{Schwarz2016}. The inclination we derive for the inner disk (22~$\pm$~6$^\circ$) is in full agreement with the inclination of the star's rotational axis derived by \cite{Broeg2007} by combining photometric monitoring and radial velocity measurements (27~$\pm$~5$^\circ$). The analysis of \cite{MacGregor2017} suggests that the outer disk inclination is significantly higher than that estimated from the star's rotation. We checked that the $\chi^2$ of the fit of our interferometric data is twice higher, when fixing, for the inner disk, the inclination and position angles of the outer disk. 

\vskip 0.05cm
\textbf{V2062~Oph (\#12)}.
It hosts a transitional disk with a compact inner disk accreting onto the central star, a large dust and gas gap (30 au-wide), and an outer ring. The inner and the outer disks are suspected to be misaligned since two symmetric shadows in the azimuthal brightness profile of the outer ring are observed in the H-band images in scattered light \citep{Casassus2018}. The inclination for the inner disk we derive from the GRAVITY measurements (32~$\pm$~4$^\circ$) perfectly matches with the stellar inclination derived from the rotation period and the $v \sin i$ (30~$\pm$~5$^\circ$) determined by \cite{Bouvier2020b}. These measurements also fully agree with the misalignment predictions made by \cite{Casassus2018} to explain the shadows observed in the environment of this star, as previously mentioned in \cite{Bouvier2020a}.

\vskip 0.05cm
\textbf{RY~Lup (\#8)}.
It has also a transitional disk, whose dusty disk has been modeled through radiative transfer by our team in \cite{Bouarour2020}. Our disk model corresponds to an inclination of the inner disk of 50$^\circ$ and we show that increasing the inclination of the inner disk to 70$^\circ$ (i.e., the inclination of the outer disk observed by SPHERE) would lead to a stellar mass in disagreement with the dynamical mass estimate from the observed rotation of the gas disk, and would not permit to correctly reproduce the GRAVITY observations. 

\vskip 0.05cm
\textbf{AS~205 (\#14)}.
{It is a multiple system located in the $\rho$-Ophiuchi star-forming region. A projected separation of 1.3" between the two components (i.e., the northern K5 PMS star AS~205~N and the southern spectroscopic binary AS~205~S) has been observed in NIR imaging \citep{McCabe2006}. \cite{Kurtovic2018} have detected two symmetric spiral arms in the disk of AS~205~N, and clear signatures of tidal interactions in the $^{12}$CO line. When fixing the ALMA inclination and position angle for the fits of our interferometric data, the $\chi^2$ increases from 3.9 up to 4.7 for the centro-symmetric ring model, and doubles for the azimuthally modulated ring model. This also agrees with the visibility differences observed at 25~M~$\lambda$ and 50~M~$\lambda$ (Fig. B.3) for baselines of same lengths but different orientations, pointing toward an inclination significantly different from pole-on.} 


Increasing the sample of protoplanetary disks for which we could compare the orientations of the inner and outer disks would be crucial to test, for transitional disks, the interpretation of the shadow features as coming from misaligned disks \citep{Marino2015ApJ...798L..44M,Min2017}, to better understand the role of close companions, and to look for trends with disk morphology and age. 






\section{Conclusions}

Optical long baseline interferometry is a unique means to directly probe the innermost regions of the dusty and gaseous environment of YSOs, down to a few 0.01~au. With the GRAVITY instrument, we resolve the compact inner dusty disks around 17 T Tauri stars, extending thus our homogeneous study of the K-band emission to 44 YSOs. With an improved sensitivity and a better spatial frequency coverage, GRAVITY allows us to revisit the pioneering works on the protoplanetary disks in the near-infrared range with a statistical view {and test various inner disk models}:
\begin{itemize}
    \item The K-band continuum emitting regions of the protoplanetary disks of the T Tauri stars appear as wide rings, like those observed in H- and K bands around the more massive Herbig Ae/Be stars. Their sizes are typically larger than those derived from the dust sublimation radius computation. 
    \item We do not detect a clear correlation between these sizes and the accretion rate of the central star. Furthermore, computing the magnetic fields producing magnetic truncation radii as large as the half-flux radii measured in the K band leads to magnetic field intensities much larger than those that have been measured in T Tauri stars. This suggests that the magnetic truncation radii are typically well within the inner gaseous disk of classical T Tauri stars, and that the dusty disk truncation might not be mainly controlled by accretion. 
    \item Our measurements agree very well with disk models including the scattered stellar light, confirming that, for cooler young stars, this scattered component should be taken into account. 
    \item When using the N-to-K size ratio as a proxy of the disk morphology and evolution, the disks with silicate features in emission seem to exhibit smaller size ratios than those with weak and/or in absorption features (i.e., with large gaps). The limited size of our sample makes it difficult to strongly confirm this trend as well as an increase in the N-to-K ratio with age. It is worth to combine measurement campaigns of K- and N-band sizes with GRAVITY and MATISSE \citep{Lopez2018} and to populate the plots of N-to-K size ratio as a function of mass and age, so as to look for a clear, universal evolution mechanism of the {inner regions of} protoplanetary disks.
    \item GRAVITY also provides us with inclinations and position angles of the inner disks. {For nine among our 17 sources, ALMA data are available. Of those nine targets, we detect clear misalignments between the inner and the outer disks as observed with ALMA for four objects, while five have good agreement between inner and outer disk, even if we do not rule out slight misalignments.} 
\end{itemize}
These observations at high angular resolution are crucial to better understand the complexity of the morphology and the dynamics of these protoplanetary disks at various spatial scales, and the presence of companions.

\bibliographystyle{aa} 
\bibliography{reference.bib}

\begin{acknowledgements}
This work was supported by CNRS/INSU, by the "Programme National de Physique Stellaire" (PNPS) of CNRS/INSU co-funded by CEA and CNES, and by Action Spécifique ASHRA of CNRS/INSU co-funded by CNES. This project has received funding from the European Research Council (ERC) under the European Union’s Horizon 2020 research and innovation programme (grant agreement No 742095; SPIDI: StarPlanets-Inner Disk-Interactions, http://www.spidi-eu.org, and grant agreement No 740651 NewWorlds). This work has made use of data from the European Space Agency (ESA) mission {\it Gaia} (\url{https://www.cosmos.esa.int/gaia}), processed by the {\it Gaia} Data Processing and Analysis Consortium (DPAC,\url{https://www.cosmos.esa.int/web/gaia/dpac/consortium}). Funding for the DPAC has been provided by national institutions, in particular the institutions participating in the {\it Gaia} Multilateral Agreement. A.C.G. has received funding from the European Research Council (ERC) under the European Union’s Horizon 2020 research and innovation program (grant agreement No. 743029). R.G.L. and Y.-I. Bouarour acknowledge support by Science Foundation Ireland under Grant No. 18/SIRG/5597. T.H. acknowledges the support of the ERC grant 832428 "Origins". A.A. and P.G. were supported by Funda\c{c}\~{a}o para a Ci\^encia e a Tecnologia, with grants reference UIDB/00099/2020, SFRH/BSAB/142940/2018 and PTDC/FIS-AST/7002/2020.

\end{acknowledgements}

\begin{appendix}

\section{Log of GRAVITY observations}

The log of the observations is given in Table~A.1.

\begin{table}[h]
\label{tab:obs}
\caption{Observation log of VLTI/GRAVITY observations. "N" denotes the number of 5-minute-long files that have been recorded on the target. Bold names denote targets for which the hydrogen line Br$\gamma$ is in emission in the GRAVITY spectrum.}
\centering
\vspace{0.1cm}
\begin{tabular}{c c c c c}
\hline
Target name & Date & Configuration & N & Calibrator \\ 
\hline
{\bf DG Tau} & 2019-01-13 & D0-G2-J3-K0 & 10 & HD~31464\\
& 2019-01-21 & UT1-2-3-4 & 10 & HD~31464\\ 
\hline
{\bf DR Tau} & 2019-01-12 & D0-G2-3-K0 & 10 & HD~31464\\
 \hline
RY Tau & 2017-12-10 & UT1-2-3-4 & 12 & HD~58923\\
\hline
{\bf T Tau N} & 2017-12-10 & UT1-2-3-4 & 6 & HD~58923\\
\hline
GQ Lup & 2019-05-24 & A0-G1-J2-J3 & 6 & HD~141977\\
& 2019-07-11 &  D0-G2-J3-K0 & 3 & HD~142383\\
\hline
IM Lup & 2019-07-13 & D0-G2-J3-K0 & 4 & HD~141922\\
\hline
{\bf RU Lup} & 2018-04-27 & UT1-2-3-4 & 7 & HD~99264\\
& 2019-05-24 & A0-G1-J2-J3 & 3 & HD~141722\\
\hline
RY Lup & 2017-06-11 & UT1-2-3-4 & 6 & HD~110878\\
\hline
{\bf S CrA N} & 2016-07-20 & UT1-2-3-4 & 2 & HD~188787 \\
& 2016-08-16 & UT1-2-3-4 & 2 & HD~176047 \\
\hline
{\bf S CrA S} & 2016-07-20 & UT1-2-3-4 & 7 & HD~188787\\
\hline
{\bf VV CrA SW} & 2019-06-20 & UT1-2-3-4 & 4 & HD~162926\\
& 2019-07-13 & D0-G2-J3-K0 & 10 & HD~186419\\
\hline
{\bf V2062 Oph} & 2019-06-22 & UT1-2-3-4 & 26 & HD~147578\\
&&&&HD~149562\\
\hline
V2129 Oph & 2018-06-27 & UT1-2-3-4 & 16 & HD~147701\\
\hline
{\bf AS 205 N} & 2018-05-18 & A0-G1-J1-K0 & 5 & HD~166165\\
& 2018-05-19 & A0-G1-J1-K0 & 10 & HD~166165\\
& 2019-07-13 & D0-G2-J3-K0 & 8 & HD~143902\\
\hline
{\bf AS 353} & 2019-04-20 & UT1-2-3-4 & 12 & HD~183442 \\
\hline 
{\bf DI Cha} & 2018-03-05 & A0-G1-J2-J3 & 8 & HD~91375\\
\hline
{\bf TW Hya} & 2019-01-21 & UT1-2-3-4 & 13 & HD~95470\\
\hline
\end{tabular}
\end{table}

\section{GRAVITY calibrated data}
\clearpage

\begin{figure*}[h]
    \centering
    \includegraphics[width=15cm]{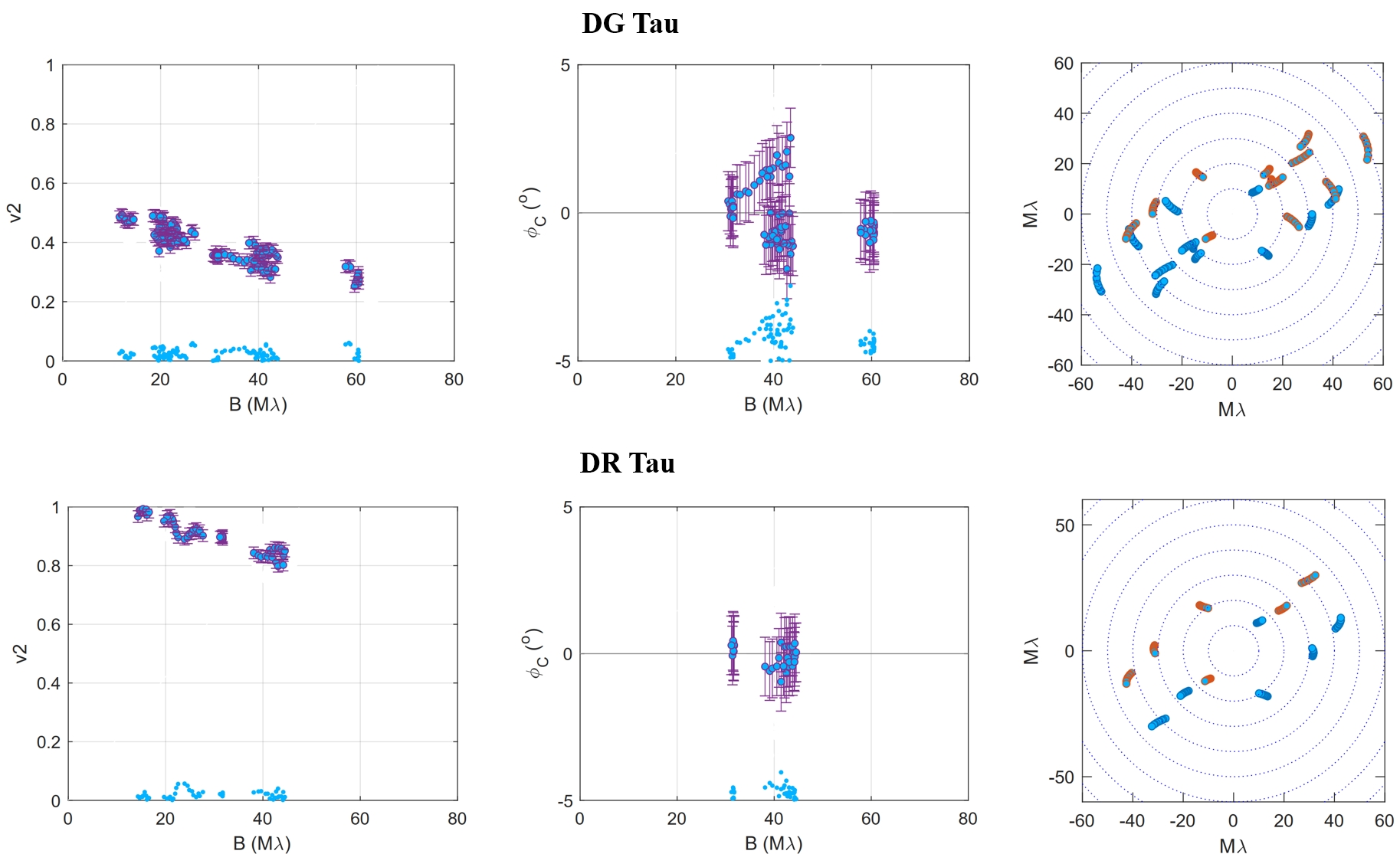}
    \includegraphics[width=15cm]{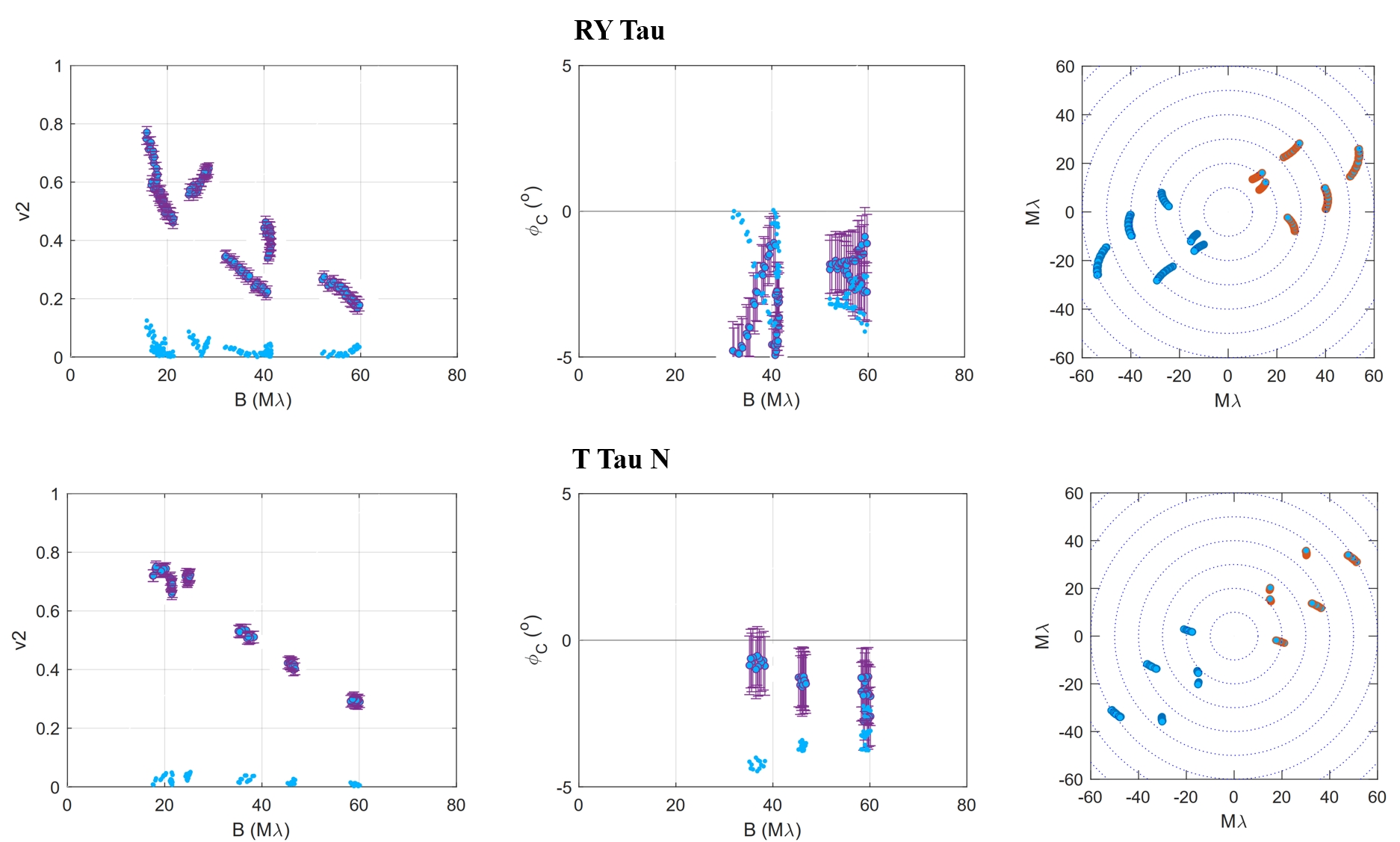}  
    \includegraphics[width=15cm]{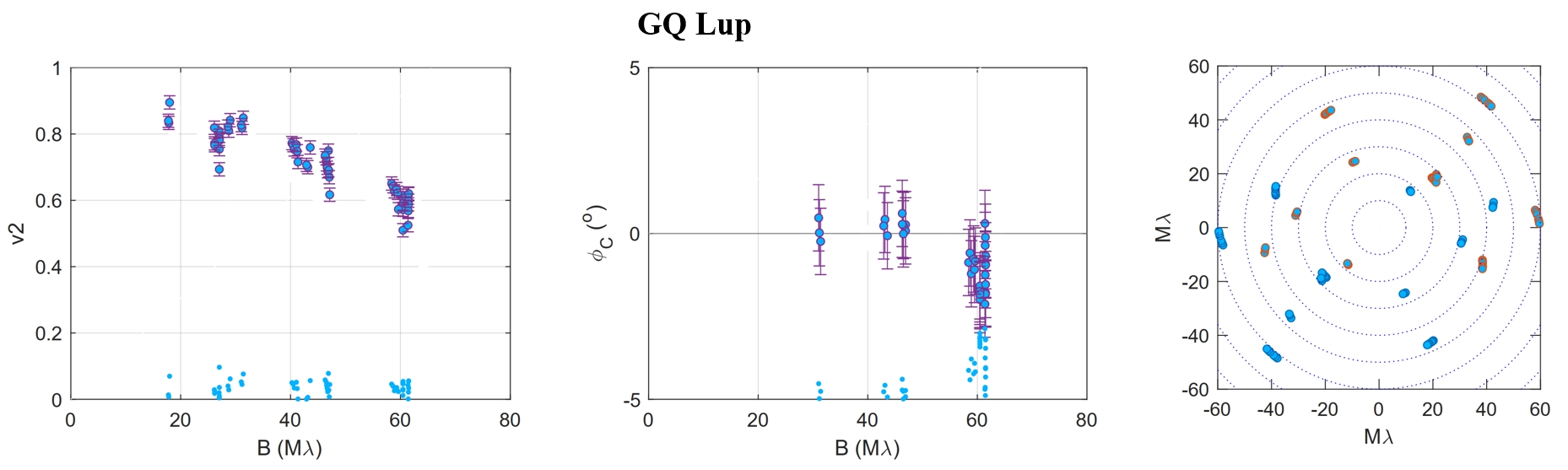}
    \caption{GRAVITY observations for the central spectral channel of the Fringe Tracker ($\lambda$~=~2.15~$\mu$m). {\bf Left.} Visibilities squared. {\bf Middle.} Closure phases. The blue symbols at the bottom of the visibility and closure phase curves display the absolute values of differences between the observations and the best-fit ring models without azimuthal modulation. These residuals are shifted by -5$^\circ$ for the closure phase plots. {\bf Right.} {\it (u, v)} planes: (u, v) coordinates in blue, (-u, -v) coordinates in red, and 10, 20, 30, 40, 50, and 60 M$\lambda$ isocontours in dotted lines.}
        \label{fig:data}
\end{figure*}

\begin{figure*}[h]
    \centering    
    \includegraphics[width=15cm]{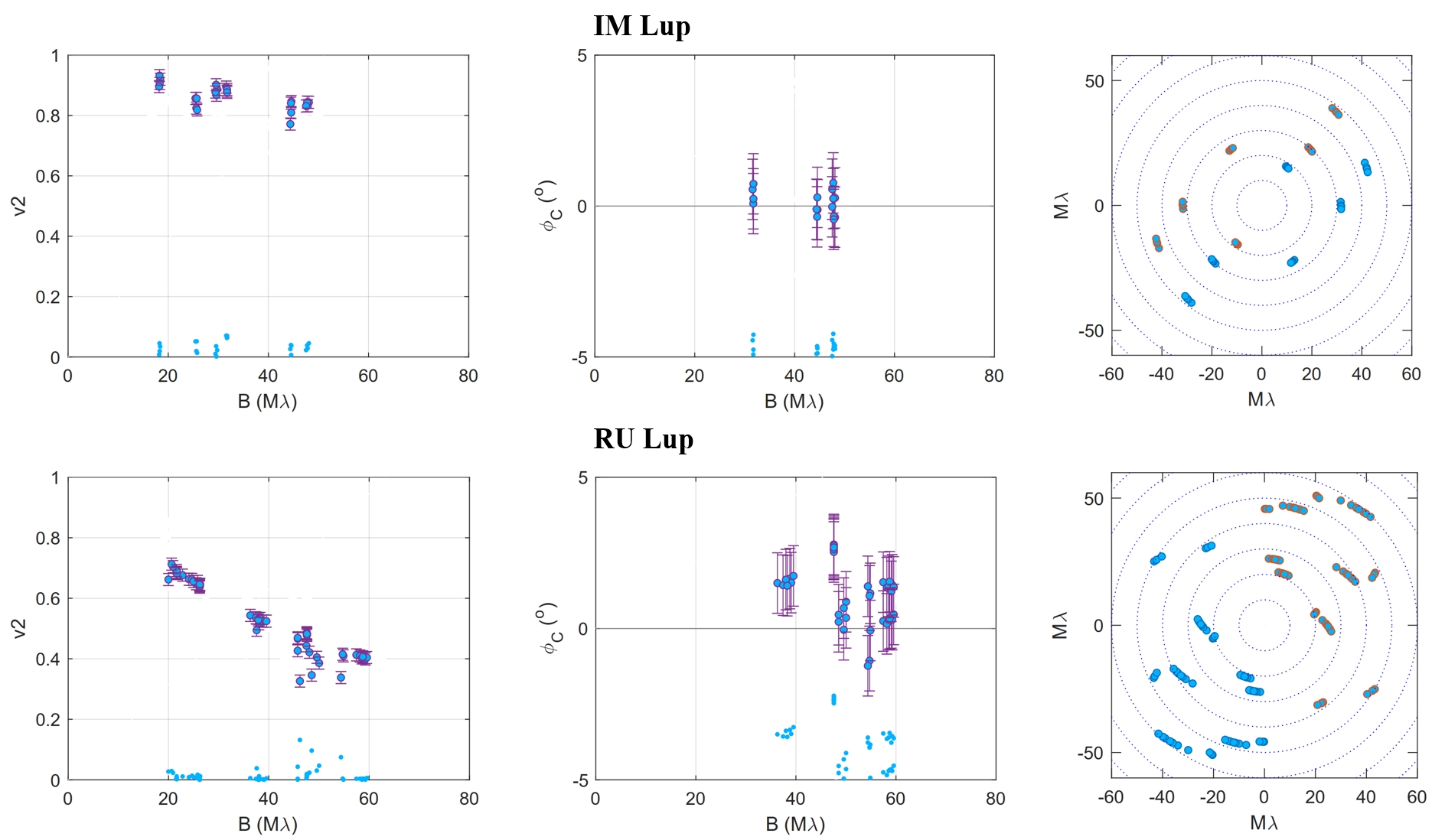}
    \includegraphics[width=15cm]{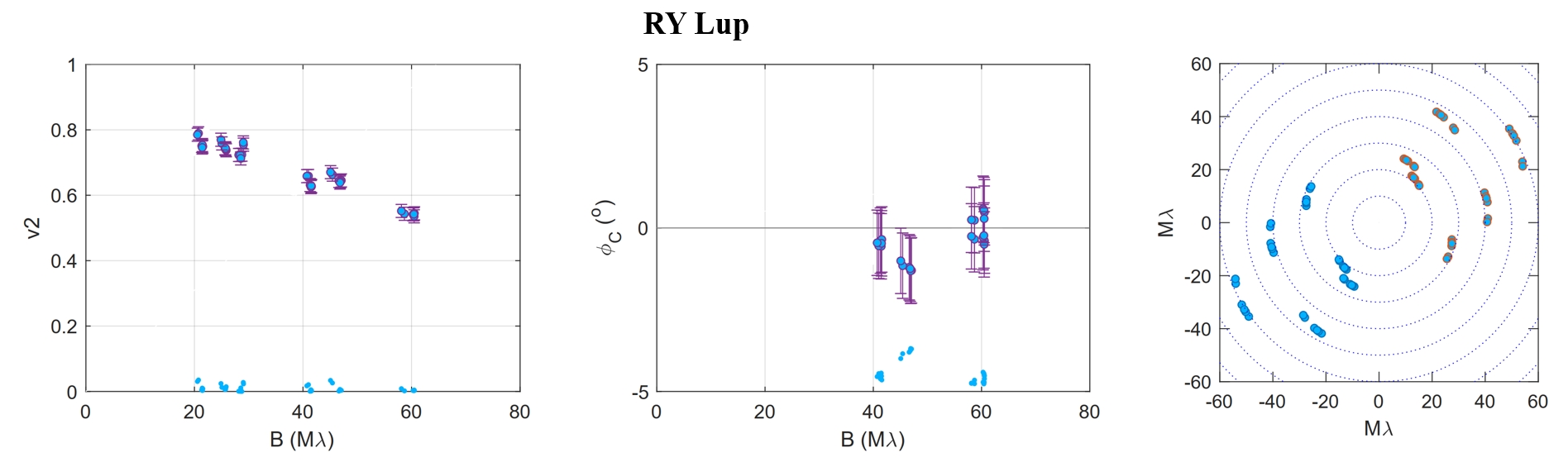}
    \includegraphics[width=15.cm]{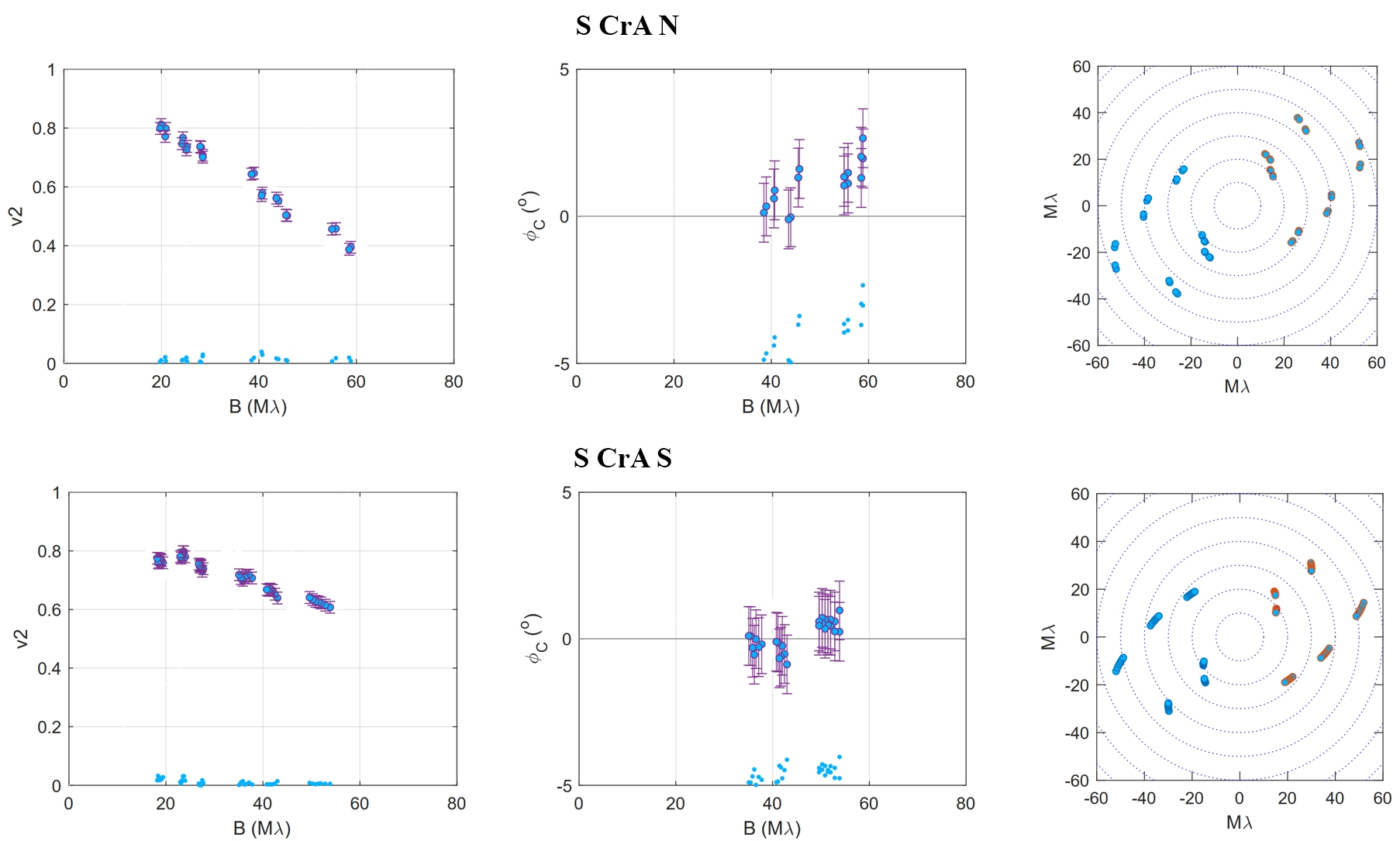}    \caption{GRAVITY observations (continued).}
    \label{fig:data}
\end{figure*}

\begin{figure*}[h]
    \centering
    \includegraphics[width=15cm]{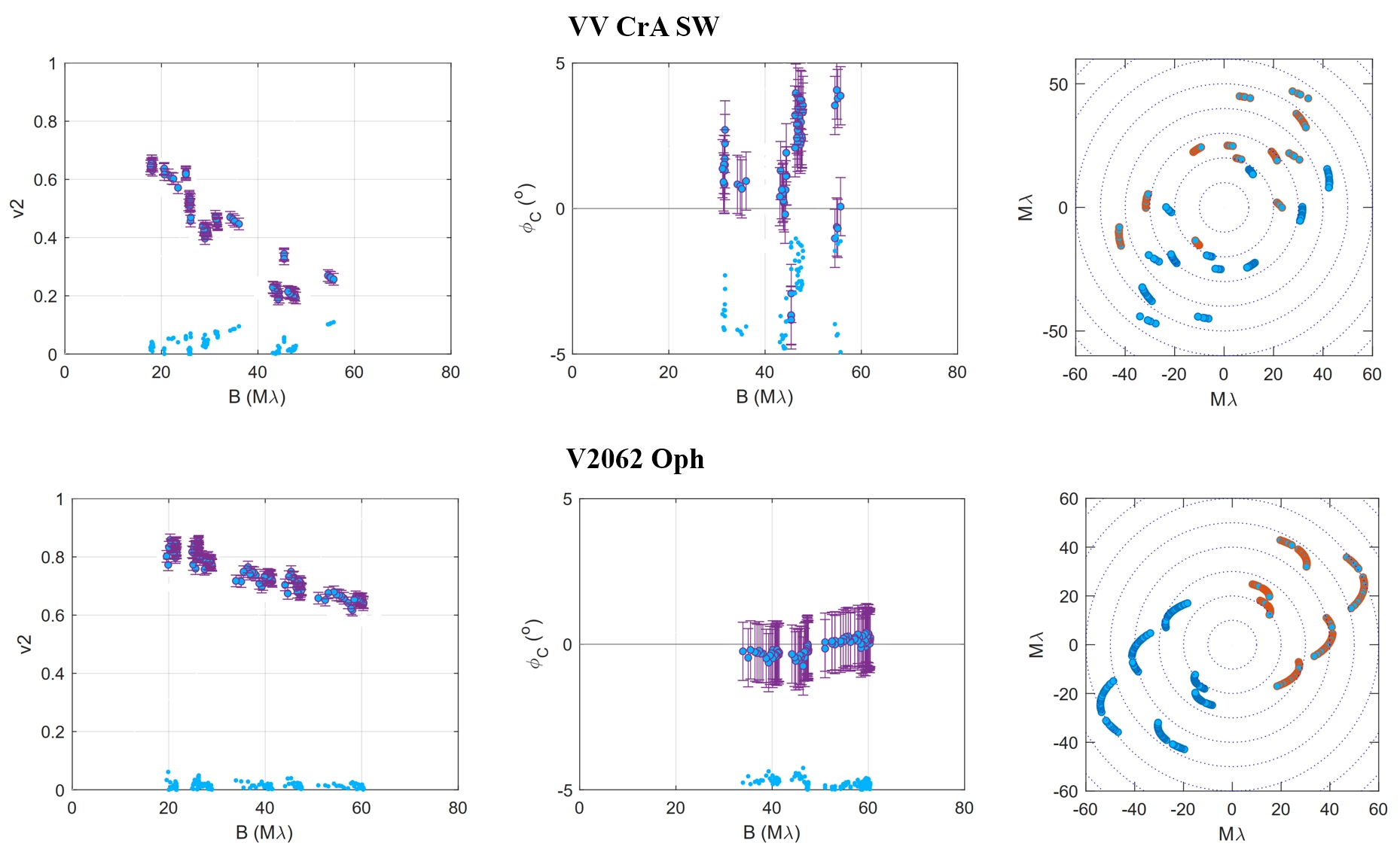}
    \includegraphics[width=15cm]{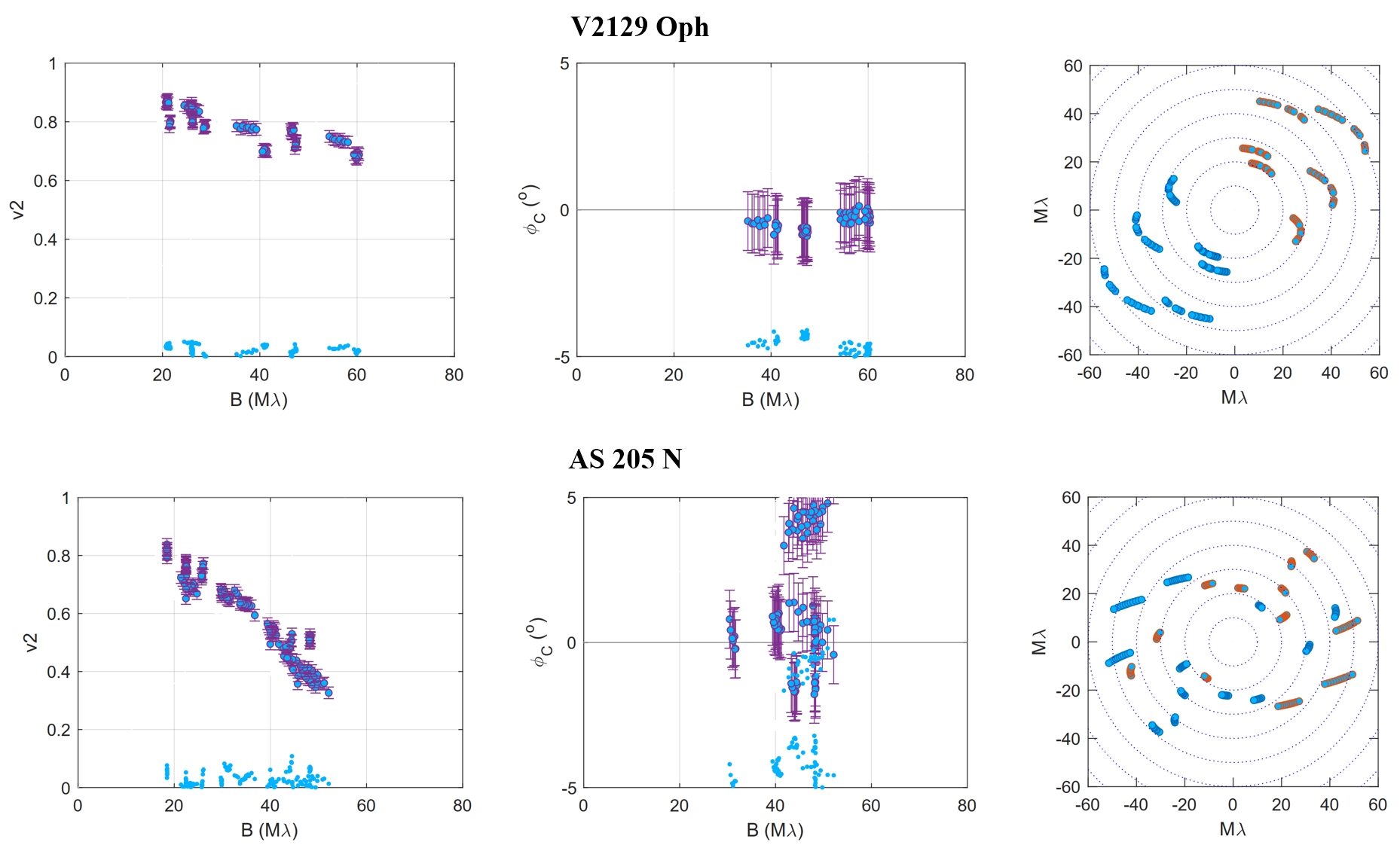}
    \includegraphics[width=15cm]{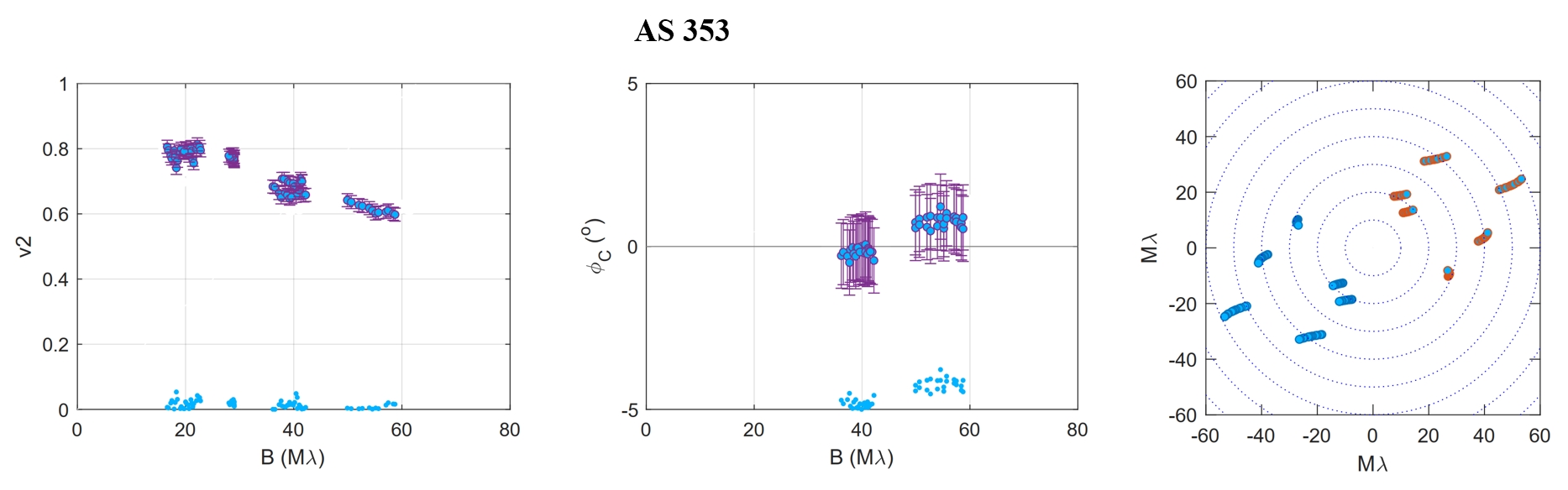}
    \caption{GRAVITY observations (continued).}
    \label{fig:data}
\end{figure*}

\begin{figure*}[h]
    \centering
    \includegraphics[width=15cm]{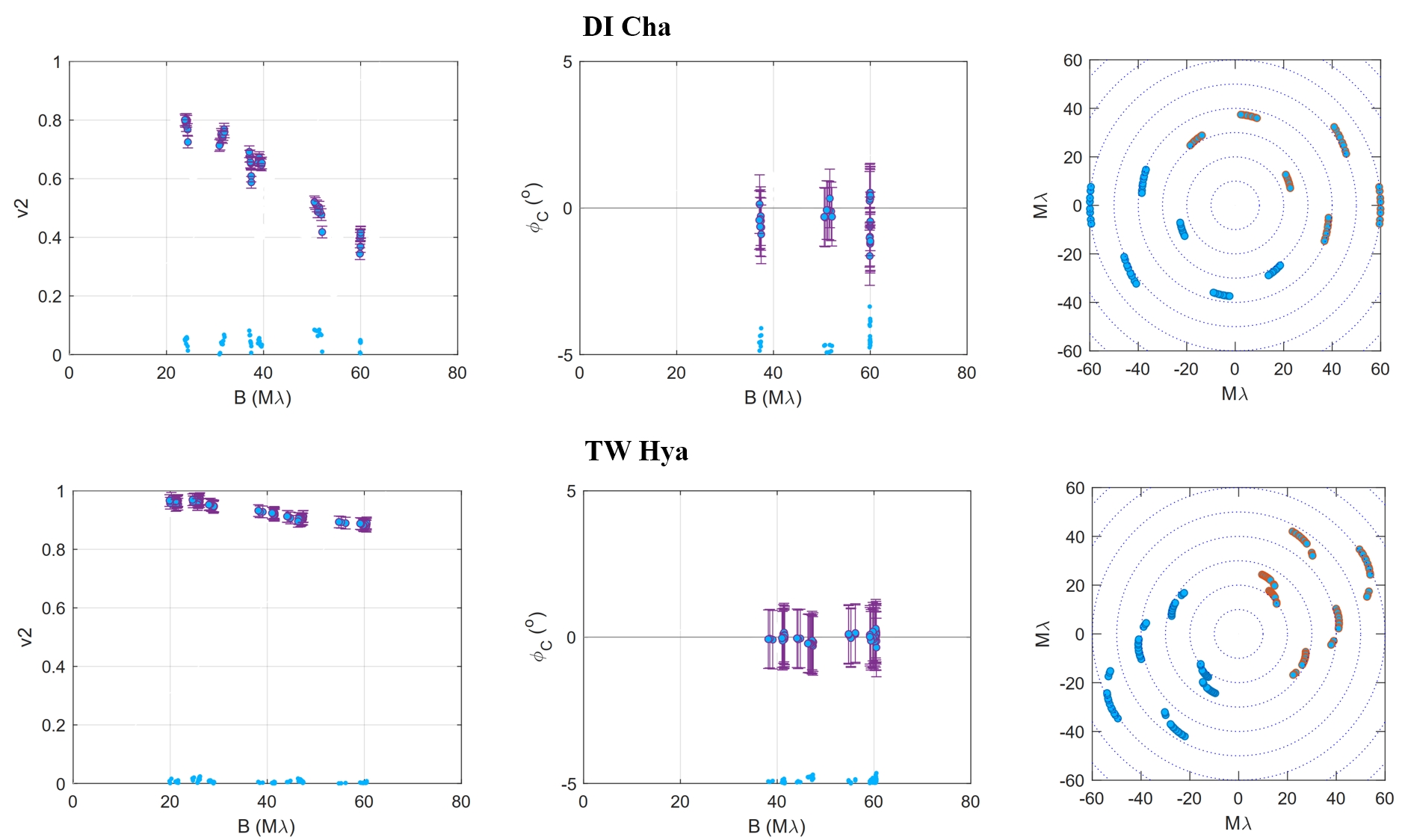}
    \caption{GRAVITY observations (continued).}
    \label{fig:data}
\end{figure*}

\clearpage

\section{Best-fit parameters for the azimuthally modulated ring model}

For the targets that exhibit closure phase signals that are non consistent with zero, we performed a fit of the interferometric data with the azimuthally modulated Gaussian ring model. The corresponding parameters are given in Table~\ref{tab:fits_ring_modulated}. 

\begin{table*}[h]
        \caption{Best-fit parameters for the azimuthally modulated Gaussian ring model with 1$\sigma$ error bars. The $\chi^2_r$ computation includes both visibilities and closure phases. 
        }
    \centering
    \begin{tabular}{cccccccccccc}
    \hline
    {\tiny Name} & {\tiny $\cos i$} & {\tiny $i$} & {\tiny PA} & {\tiny $f_c$}  & {\tiny $f_h$}  & \multicolumn{2}{c}{\tiny Half-flux radius} & {\tiny $w$}& {\tiny $c_1$} & {\tiny $s_1$} & {\tiny $\chi^2_r$} \\ 
    {\tiny --} & {\tiny --} & {\tiny [$^\circ$]} & {\tiny [$^\circ$]} & {\tiny [\%]} & {\tiny [\%]} & {\tiny [mas]} & {\tiny [au]} & {\tiny --} & {\tiny --} & {\tiny --} & {\tiny --} \\
    \hline
   {\tiny RY Tau} & {\tiny 0.51~$\pm$~0.02} & {\tiny 59~$\pm$~2} & {\tiny 7~$\pm$~2} & {\tiny 50~$\pm$~5} & {\tiny 4~$\pm$~2} & {\tiny 2.69$^{+0.14}_{-0.18}$} & {\tiny 0.37~$\pm$~0.02} & {\tiny 0.99~$\pm$~2.01} & {\tiny -0.65~$\pm$~0.15} & {\tiny 0.70~$\pm$~0.15} & {\tiny 3.2}  \\
   & \\
    {\tiny VV CrA SW} & {\tiny 0.83~$\pm$~0.03} & {\tiny 34~$\pm$~3} & {\tiny 113~$\pm$~4} & {\tiny 85~$\pm$~5} & {\tiny 14~$\pm$~2} & {\tiny 1.15~$\pm$~0.05} &  {\tiny 0.180~$\pm$~0.008} & {\tiny 0.59~$\pm$~0.07} & {\tiny -0.65~$\pm$~0.14} & {\tiny 0.40~$\pm$~0.08} & {\tiny 4.5}  \\
    & \\
  {\tiny AS 205 N} & {\tiny 0.65~$\pm$~0.02} & {\tiny 49~$\pm$~2} & {\tiny 90~$\pm$~1} & {\tiny 45~$\pm$~5} & {\tiny 8~$\pm$~2} & {\tiny 1.40$^{+0.11}_{-0.09}$} &  {\tiny 0.18$^{+0.015}_{-0.012}$} & {\tiny 0.3~$\pm$~0.18} & {\tiny 1.00~$\pm$~0.01}& {\tiny 0.04~$\pm$~0.04} & {\tiny 1.9} \\
  & \\      \hline
    \end{tabular}
    \label{tab:fits_ring_modulated}
\end{table*}

\section{Characteristic sizes}

The characteristic sizes of our targets are gathered in Table~\ref{tab:charac_sizes}.

\begin{table*}[h]
\label{tab:Prot}
\caption{Rotation periods, $v \sin i$, corotation radii, magnetic truncation radii for a magnetic field of 1~kG, and sublimation radii as computed from Eq. (2) with $T_{\rm sub}$~=~1500~K for our sample. The last column gives the references for the rotational period or the $v \sin i$.}
\centering
\vspace{0.1cm}
\begin{tabular}{c c c c c c c}
\hline
Target name & $v \sin i$ [km/s] & $P_{\rm rot}$ [days] & $R_{\rm co}$ [au] & $R_{\rm mag}$ [au] & $R_{\rm sub}$ [au] & References                  \\ 
\hline
DG Tau &  & 6.3 & 0.06 & 0.016 & 0.05 -- 0.10 & \cite{Bouvier1993}\\
DR Tau &  & 5.0 & 0.07 & 0.045 & 0.07 -- 0.14 & \cite{Percy2010} \\
RY Tau &  & 5.6 & 0.09 & 0.114 & 0.15 -- 0.37 & \cite{Herbst1987}\\
T Tau N & & 2.8 & 0.06 & 0.060 & 0.11 -- 0.26 & \cite{Herbst1986}\\
GQ Lup &  & 8.4 & 0.07 & 0.044 & 0.04 -- 0.08 & \cite{Donati2012}\\
IM Lup &  & 7.2 & 0.06 & 0.12 & 0.05 -- 0.11 & from TESS (unpublished)\\
RU Lup &  & 3.7 & 0.04 & 0.032 & 0.05 -- 0.10 & \cite{Covino1989}\\
RY Lup &  & 3.9 & 0.06 & 0.036 & 0.06 -- 0.14 & \cite{Bouvier1986}\\
S CrA N &  & 3.6 & 0.04 & 0.047 & 0.06 -- 0.11 & Nowacki et al. (private comm.)\\
S CrA S & 12 & 3 & 0.04 & 0.028 & 0.04 -- 0.07 & \cite{MaccVVCrA}\\
VV CrA SW & 3.6 & 11 & 0.09 & 0.046 & 0.04 -- 0.08 & \cite{MaccVVCrA}\\
V2062 Oph &  & 2.96 & 0.05 & 0.030 & 0.04 -- 0.10 & \cite{Bouvier2020b}\\
V2129 Oph &  & 6.53 & 0.07 & 0.094 & 0.06 -- 0.13 & \cite{MaccV2129Oph}\\
AS 205 N &  & 2.7/6.78 & 0.04/0.07 & 0.012 & 0.08 -- 0.20 & \citet{Covino1992}/\cite{Artemenko2010}\\
AS 353 & 10 & 7 & 0.1 & 0.044 & 0.05 -- 0.10 & \cite{MaccAS353}\\
DI Cha & 30 & 1.5 & 0.04 & 0.056 & 0.14 -- 0.40 & \cite{Bouvier1986}\\
TW Hya &  & 3.56 & 0.05 & 0.023 & 0.02 -- 0.04 & \cite{Donati2011}\\
\hline
\end{tabular}
    \label{tab:charac_sizes}
\end{table*}

\section{Comparison with previous size determinations}

\begin{figure}[h]
    \centering
    \includegraphics[width=7.5cm]{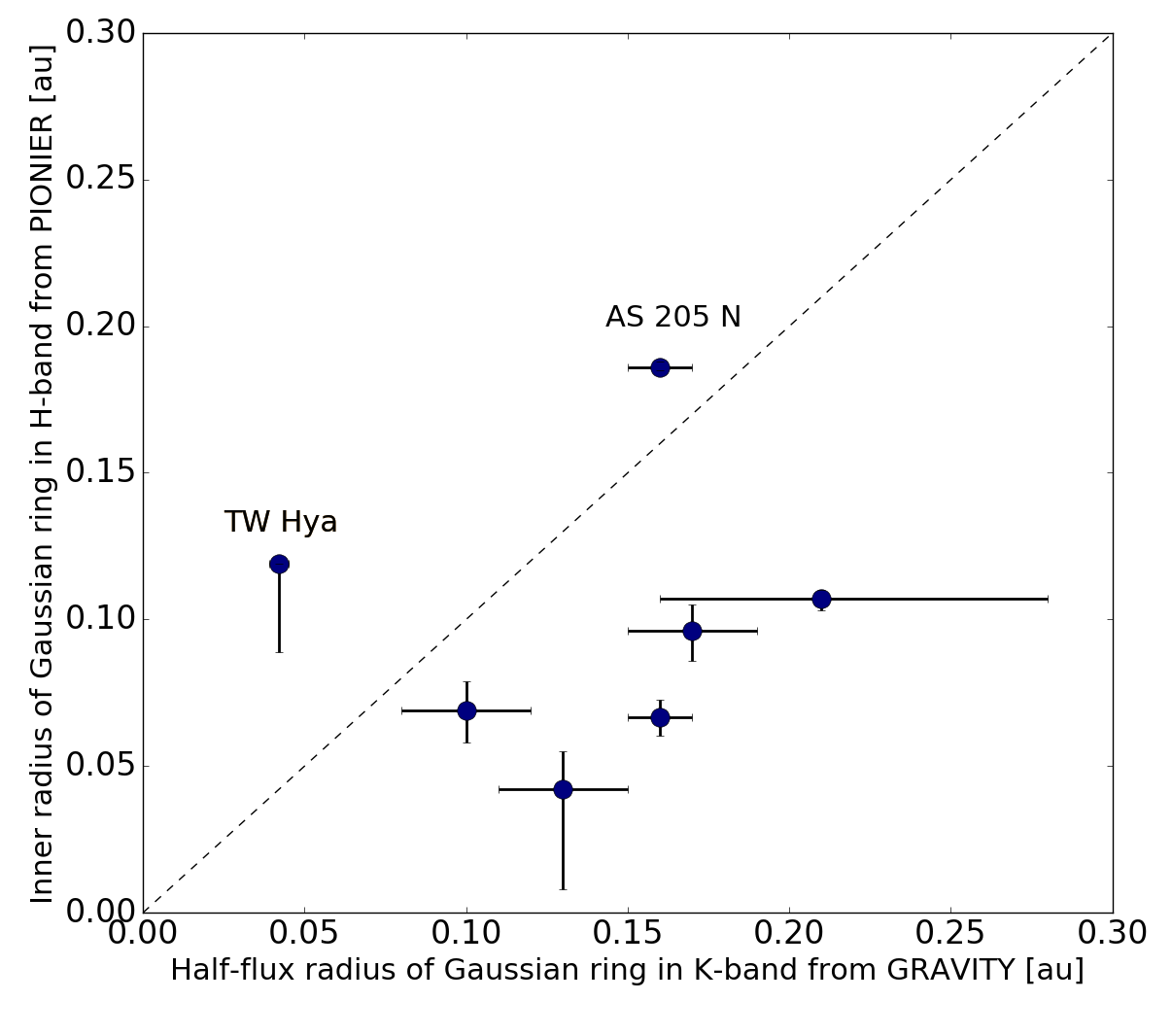}
    \caption{Inner radii of the Gaussian ring in the H band derived from PIONIER as a function of the half-flux radii in the K band derived from GRAVITY interferometric measurement. The dashed line shows the 1:1 relation.}
    \label{fig:PIONIER}
\end{figure}

Fig.~\ref{fig:PIONIER} presents the comparison between the K-band GRAVITY half-flux radii and the H-band sizes derived from PIONIER measurements whose values are given in Table~\ref{tab:PIONIER}. 

\begin{table}[h]
\centering
\caption{H-band inner radii derived from PIONIER \citep{Anthonioz2015} and corrected for Gaia EDR3 distances and $\chi^2$ of the PIONIER fits.}
\begin{tabular}{ccc}
\hline
 Object & $R_{in}$ [au] & $\chi^2$ \\
\hline
 GQ Lup  &  0.042$^{+0.013}_{-0.034}$ & 1.36     \\ [1ex]
 RU~Lup  & 0.107$^{+0.003}_{-0.004}$ & 2.20  \\[1ex]
 RY~Lup  & 0.067$\pm$0.006 & 3.83 \\[1ex]
 S~CrA~N  & 0.096$^{+0.009}_{-0.010}$ & 0.69 \\[1ex]
 V2129~Oph   &  0.069$^{+0.010}_{-0.011}$ & 2.55\\[1ex]
 AS~205~N  & 0.186$^{+0.002}_{-0.001}$ & 13.79  \\[1ex]
 TW~Hya &  0.12$^{+0.03}_{-0.00}$ & 0.89  \\[1ex]
\hline
\end{tabular}
\label{tab:PIONIER}
\end{table}

\begin{figure}[h]
    \centering
    \includegraphics[width=7.5cm]{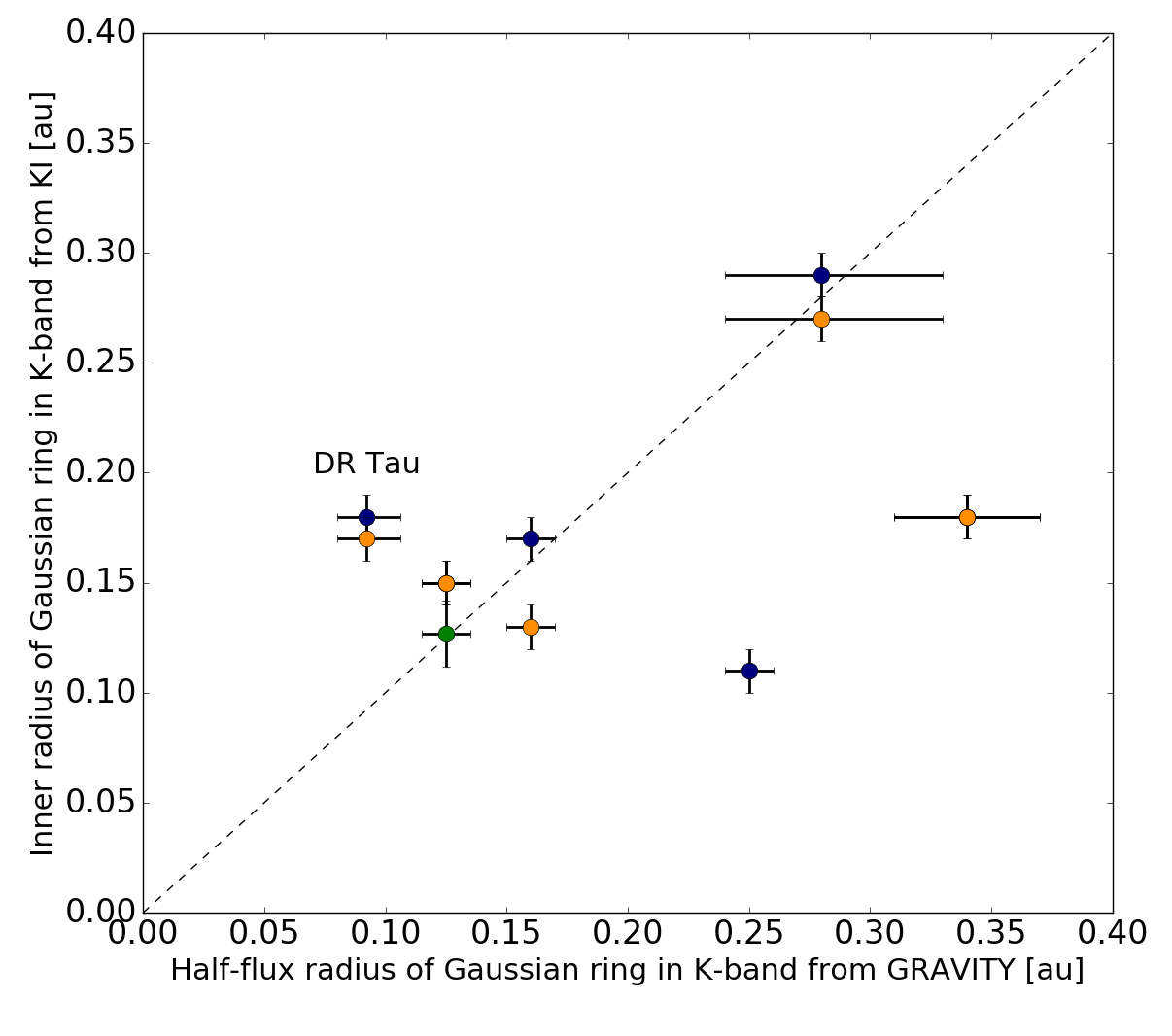}
    \caption{Inner radii of the ring in the K band derived from the KI by \cite{Eisner2010} (orange symbols), by \cite{Eisner2014} (blue symbols), and by \cite{Akeson2005} (green symbol) as a function of the half-flux radii in the K band derived from GRAVITY. The dashed line shows the 1:1 relation.}
    \label{fig:KI}
\end{figure}

Fig.~\ref{fig:KI} displays the comparison between the K-band GRAVITY half-flux radii and the K-band sizes derived from KI measurements whose values are gathered in Table~\ref{tab:KI}.

\begin{table}[h]
\centering
\caption{K-band angular diameters derived from the KI and corresponding inner radii in au corrected for Gaia EDR3 distances.}
\begin{tabular}{cccc}
\hline
 Object & Angular diameter [mas] & $R_{in}$ [au] & Refs \\
\hline
 DG~Tau  & 2.44~$\pm$~0.02  & 0.15$\pm$0.01   & (a)   \\
  & 2.40~$\pm$~0.03  & 0.15$\pm$0.01   &  (b) \\
  & 2.03$^{+0.45}_{-0.48}$ & 0.127$^{+0.014}_{-0.015}$& (c) \\[1ex]
 DR~Tau  & 1.72~$\pm$~0.04  & 0.17$\pm$0.01   & (a, b) \\[1ex]
 RY~Tau  & 2.63~$\pm$~0.03  & 0.18$\pm$0.01   & (a) \\
  & 2.57~$\pm$~0.02  & 0.18$\pm$0.01   & (b)  \\[1ex]
T~Tau~N   & 1.54~$\pm$~0.07  & 0.11$\pm$0.01   &  (b) \\[1ex]
 AS~205~N  & 2.02~$\pm$~0.07  & 0.13$\pm$0.01   & (a) \\
  & 2.52~$\pm$~0.04  & 0.17$\pm$0.01   & (b) \\[1ex]
   AS~353  & 1.36~$\pm$~0.06  & 0.27$\pm$0.01   & (a)  \\
  & 1.47~$\pm$~0.06  & 0.29$\pm$0.01   &(b) \\
\hline
\end{tabular}\label{tab:KI}
    \tablefoot{\noindent ${(a)}$ \cite{Eisner2010}; ${(b)}$ \cite{Eisner2014}; ${(c)}$ \cite{Akeson2005}.}
\end{table}


\end{appendix}

\end{document}